\documentclass[lettersize,journal]{IEEEtran}
\usepackage{amsmath,amsfonts}
\usepackage{algorithmic}
\usepackage{algorithm}
\usepackage{array}
\usepackage{textcomp}
\usepackage{stfloats}
\usepackage{multirow}
\usepackage{xcolor,colortbl}
\usepackage[hyphens]{url} 
\usepackage{lipsum}
\usepackage{verbatim}
\usepackage{graphicx}
\usepackage{cite}
\usepackage{xcolor}
\usepackage{float}
\usepackage{caption}
\usepackage{subcaption}

\newcommand{\backup}[1]{} 

\definecolor{Gray}{gray}{0.85}
\definecolor{LightCyan}{rgb}{0.88,1,1}

\newcolumntype{a}{>{\columncolor{LightCyan}}c}
\newcolumntype{b}{>{\columncolor{white}}c}

\begin{document}


\title{Open RAN Testbeds with Controlled Air Mobility}

\author{Magreth Mushi, Yuchen Liu,~\IEEEmembership{Member,~IEEE}, Shreyas Sreenivasa, Ozgur Ozdemir, Ismail Guvenc,~\IEEEmembership{Fellow,~IEEE}, Mihail Sichitiu,~\IEEEmembership{Member,~IEEE}, Rudra Dutta,~\IEEEmembership{Senior~Member,~IEEE}, and Russ Gyurek
\thanks{This work was supported in part by the NSF Award CNS-1939334. M. Mushi, Y. Liu, R. Dutta are with the Dept. Computer Science, NC State University, Raleigh, NC; S. Sreenivasa, O. Ozdemir, I. Guvenc, M. Sichitiu are with the Dept. Electrical and Computer Engineering, NC State University, Raleigh, NC; Russ Gyurek is with Cisco Systems, Raleigh, NC.}
}




\maketitle

\begin{abstract}
With its promise of increasing softwarization, improving disaggregability, and creating an open-source based ecosystem in the area of Radio Access Networks, the idea of Open RAN has generated rising interest in the community.  Even as the community races to provide and verify complete Open RAN systems, the importance of verification of systems based on Open RAN under real-world conditions has become clear, and testbed facilities for general use have been envisioned, in addition to private testing facilities.  Aerial robots, including autonomous ones, are among the increasingly important and interesting clients of RAN systems, but also present a challenge for testbeds.  Based on our experience in architecting and operating an advanced wireless testbed with aerial robots as a primary citizen, we present considerations relevant to the design of Open RAN testbeds, with particular attention to making such a testbed capable of controlled experimentation with aerial clients. We also present representative results from the NSF AERPAW testbed on Open RAN slicing, programmable vehicles, and programmable radios. 
\end{abstract}


\section{Introduction} \label{introduction}

Open Radio Access Network (specifications defined in the O-RAN Alliance) has emerged as a serious and perhaps critically necessary alternative to the proprietary radio access network (RAN) solutions that have characterized cellular networks. In particular, Open RAN provides a richer eco-system based on the virtualization of network functions providing greater economies of scale and reduced cost. The open architecture of Open RAN, and the definition of interfaces among modules that have been thus far treated as essentially monolithic, are expected to ensure inter-operation between products from different providers, and a competitive market, leading to improved quality and lower cost of ownership.  It also enables the inclusion of commodity controllers, and the ability of operators to develop their custom control applications on top of those controllers, bringing the power of software-defined networking to RANs on an open-interface basis.

Such disaggregation comes at the cost of increased overhead, and early Open RAN systems are widely expected to have higher overheads and lower efficiency compared to extant single-vendor systems that, after all, have evolved and been integrated for decades.  Optimistic views consist of expectations of workable, if inefficient, implementations soon, followed by rapid improvements in performance.  Pessimistic views incline to doubts regarding how long such a process might take, or whether such systems can approach the efficiency of proprietary monolithic systems, or even be workable at scale.  However, there are significant gains in terms of economies of scale through virtualization as well as additional functionality that provides a much richer set of capabilities (e.g., RIC apps, etc). 

To dispassionately and pragmatically assess the workability of Open RAN, the community must move beyond early experiments and greenfield deployments to demonstrable repeatability of predictable system performance.  Designing dependable test facilities for Open RAN components and systems, therefore, is among the most important outstanding tasks of the Open RAN community at this time. 
A key promise of Open RAN is interoperability (multi-vendor), and the key to verifying such claims is through interoperability testing (IOT). Recognizing the importance of IOT, the O-RAN Alliance has dedicated two entire work groups (WG4 and WG5) to specifying interfaces, and both groups have published specifications on interoperability testing and profiles in addition to unit test specifications (see~\cite{ORAN-WG4-IOT-spec,ORAN-WG5-IOT-spec} and other specifications of WG4 and WG5).  Such profiles allow the interoperability of any set of components to be tested in test configurations that can be realized in lab-environment test benches. However, to engender the above-mentioned growing confidence, Open RAN ecosystem players (contributors, as well as vendors, operators, and users) need to be able to test components in a comprehensive end-to-end test facility - one that is embedded in a realistic setting and span in the real world, including at least in part an outdoor setting, with a non-trivial number of UEs interacting with a non-trivial number of base stations. In the rest of this paper, we reserve the term ``testbed'' to indicate facilities capable of such complete RAN system tests.

Unpeopled Aerial Vehicles (UAVs), 
have long been generally acknowledged as important clients of any future wide-area wireless communications system.  However, the full scope of such devices as denizens of the wireless communication world is only coming to be appreciated recently.  A key observation is that UAVs are not only wireless communications clients for command-and-control (the most obvious use case), but play roles in at least two other ways in the wireless ecosystem.  First, trivially, as such devices increase in intelligence, and are tasked with increasingly more sophisticated missions, these missions are likely to pose additional -- and likely much heavier -- communication requirements; for example streaming live on-site video back to the cloud, or engaging in other data-heavy cloud-assisted distributed computation tasks.  More importantly, and more significantly in the current context, with increasing on-board compute intelligence, such devices are capable of engaging not just as clients, but as crucial parts of the wireless communication infrastructure itself.  This is especially important in an open interoperable ecosystem such as Open RAN aspires to be, as open competition spurs innovative contributors to explore previously unoccupied ecosystem roles.

The visioning and design exercise for an Open RAN testbed that aspires to provide interoperability and system testing capabilities, 
if such a facility expects to support the full evolutionary arc of aerial devices, must include reflection specific to these considerations.  In this paper, we leverage our joint experience in (i) architecting and operating an advanced wireless testbed with aerial robots as primary citizens, and (ii) industry Open RAN testing and dependability expectations, to provide a starting point that we hope will be useful to such architects and designers.  In the next section, we briefly review some existing test facilities with the capability (or potential near capability) of acting as Open RAN testing resources and juxtapose them with industry Open RAN testing norms, as well as basic support requirements for UAVs.  In Section~\ref{sec:usecase}, we discuss in further detail the class of use cases that represent the potential synergistic use of UAVs in Open RAN systems.  Finally, we provide a deep consideration of one extant testbed -- our own NSF AERPAW platform at NC State University -- to showcase the process of reviewing testbed capabilities to articulate both strengths and shortcomings in light of an ideal Open RAN testbed with native UAV support.

%

\section{Visioning an Open RAN/UAV Testbed}\label{requirements}


\subsection{Existing Wireless Testbeds and System Testing} \label{orantb}



There are numerous testbeds that are accessible to researchers to experiment with wireless technologies including 5G, Open RAN, and UAVs. In Table~\ref{table:testbeds}, we provide a list a few of these testbed facilities that are accessible to researchers from academia, government, and industry. 
Note that we do not intend to present Table~\ref{table:testbeds} as either comprehensive or authoritative.  There are likely many facilities that we are unaware of, or for which no information is publicly available to us.  Even for those we have surveyed, Table~\ref{table:testbeds} represents our best knowledge as obtained from publicly available sources (as cited); we regret any unintended mischaracterization.  Our survey was also heavily biased toward facilities in the USA.

Nevertheless, since our focus is on test facilities publicly or generally available to researchers and practitioners in the US, and on facilities sizeable enough for UAVs to be practically a part of the test ecosystem, we believe that Table~\ref{table:testbeds} provides representative, and meaningfully extensive, information for the Open RAN testbed designer of the near future.
We have chosen to characterize each facility listed by means of a few high-level considerations.  Obviously, explicit currently stated support of Open RAN testing, and UAV support/integration, are features we looked for.  Related to Open RAN, we also looked at the RF spectrum the facility is capable of and allowed to operate in, by noting if it lists an Innovation Zone (IZ) license from the Federal Communications Commission (see for example~\cite{FCC-IZ}), and also its deployment context (indoor facilities may be able to use isolation such as Faraday cages and operate without an FCC Innovation Zone or experimental licenses).

Related to UAV support, we also looked at whether such UAVs (or any component of the testbed, for those without UAVs) support controlled mobility.  We consider this feature an important one for future Open RAN testbeds.  A significant proportion of wireless communications system complexity arises from (or is exacerbated by) the mobility of system components, most usually that of User Equipment (UE); therefore it it important for the testbed to support  experiments involving mobility, hand-over, and disconnect-reconnect events.  However, the core of the scientific method is the repeatability of experiments and the reproduction of experimental results.  To provide this for experiments related to mobility, the relative motion of various system components must be possible to precisely reproduce on demand, for as many runs of an experiment as necessary.

Another key feature we looked for was \emph{emulation support}.  The single most valuable characteristic of actual wireless test facilities is the availability of a real Radio Frequency (RF) environment, providing real-world challenges such as fading, multi-path, and statistical uncertainty, simultaneously with the experiment repeatability.  The \emph{simulation} of RF environments by means of mathematical models, no matter how sophisticated, abstracts a measure of realism from test results; further, the experimenter has no need of an experimental facility (or even actual radios) for simulation exercises, which are an appropriate earlier stage in proving research before considering testbed validation.  The exercise of \emph{emulation}, on the other hand, provides an important added value to a testbed, in that it is a digital twin of a real RF system, capable of operating in real-time, in which physical radio equipment can actually be immersed.  Emulation systems are driven by calibration (to some real RF environment) rather than modeling and may be realized by digital twinning, or more often by analog RF circuitry.  In extreme cases, a test facility may be entirely based on emulation, as in the case of the Colosseum system (originally created by DARPA and currently operated at Northeastern University under NSF aegis; see Table~\ref{table:testbeds}).  More typically, emulation support is an adjunct part of a physical test facility that can serve as an early and less costly stage of full testbed validation.

\begin{table*}[!h]

\caption{Existing testbeds with advanced wireless and UAV experimentation capabilities. Public testbeds indicated with an asterisk (*) may be open only to partners or require contacting testbed operators rather than being generally available through an experimentation portal. Features for which public information could not be found are marked as Not Known (NK).}
    \def\arraystretch{1}
\begin{tabular}{|p{2.5cm}|p{1.9cm}|p{1.2cm}|p{1cm}|p{1cm}|p{1.2cm}|p{0.6cm}|p{1.9cm}|p{1.6cm}|p{0.8cm}|}
\hline \rowcolor{LightCyan}
\textbf{Testbeds (alphabetical)} & \textbf{Location} & \textbf{Emulation Support} & \textbf{Open RAN Support} & \textbf{UAV Support} & \textbf{Controlled Mobility} & \textbf{FCC-IZ} & \textbf{Main Focus Area} & \textbf{Deployment Environment} & \textbf{Access}\\
\hline
AERPAW~\cite{aerpaw1} & Raleigh, NC & \checkmark & Partial & \checkmark  &\checkmark  &  \checkmark & UAVs, SDRs & Rural, Urban & Public\\
\hline
ARA~\cite{zhang2022ara} & Central Iowa, IA &\checkmark & $\times$ & $\times$ & $\times$  & $\times$ & Rural wireless & Rural & Public\\
\hline
Arena~\cite{bertizzolo2020arena} & Boston, MA & \checkmark & \checkmark & $\times$ & $\times$ & \checkmark & SDRs & Indoor grid & Public* \\
\hline
ARLIS~\cite{ARLIS_umd} & College Park, MD & $\times$ & $\times$ & $\times$& $\times$& $\times$  & 5G security & Virtual & Public* \\
\hline
ARM / Tech Mahindra 5G Lab~\cite{5gsolutionlab}& NK & NK &\checkmark  & $\times$  &$\times$  &$\times$  & 5G testing & NK & Private \\
\hline
Booz Allen 5G Lab~\cite{boozAl}& Annapolis Junction, MD & NK & NK & $\times$ &  $\times$ & $\times$ & Mission critical 5G & NK & Private\\
\hline
CCI xG Testbed~\cite{CCI_xG_Testbed}& Arlington, VA& NK  & \checkmark & $\times$ & $\times$ & $\times$ & SDRs, AI & Indoor & Public* \\
\hline
Colosseum~\cite{9677430} & Burlington, MA &\checkmark &\checkmark & $\times$ & $\times$ & \checkmark & Emulation, SDRs & Cloud & Public \\
\hline
CORNET~\cite{CORNET_VT} &Blacksburg, VA & $\times$  &\checkmark & $\times$ & $\times$ &$\times$  & SDRs & Indoor, Rooftop & Public \\
\hline
COSMOS~\cite{raychaudhuri2020challenge} & Manhattan, NY& \checkmark &\checkmark & $\times$ & $\times$ & \checkmark & mmWave, backhaul & Urban & Public\\
\hline
Drexel Grid~\cite{8824901} &Philadelphia, PA & \checkmark & $\times$  &  $\times$ & $\times$ &  $\times$ & Emulation, SDRs & Indoor grid & Public* \\
\hline
Ericsson Open Lab~\cite{ericssonOl}& NK & \checkmark & \checkmark & $\times$ & $\times$ & $\times$ & CloudRAN, virtualized 5G & Indoor & Private\\
\hline
INL Wireless Testbed~\cite{INL_Testbed}& Idaho Falls, ID& $\times$ & $\times$ &\checkmark  & Partial & $\times$ & Wireless security & Rural & Private \\
\hline
IRIS~\cite{8593636} & Los Angeles, CA & $\times$& $\times$ & $\times$ & \checkmark  & $\times$ & Robotic wireless networks & Indoor & Public*\\
\hline
LinQuest Labs~\cite{linqlab}& Chantilly, VA & \checkmark & NK & \checkmark & NK & $\times$ & 5G security, UAV, NTN & Cloud, indoor & Public*\\
\hline
NASA MTBs~\cite{nasarotor} & $\times$ &$\times$ &$\times$  &\checkmark  & $\times$ & $\times$ & Multirotor UAV testing & Indoor & Public*\\
\hline
New York UAS Test Site~\cite{nytestsite} & Rome, NY& $\times$ & $\times$ & \checkmark & Partial & $\times$ & BVLOS UAV testing & Rural, Urban  & Public*\\
\hline
NIST 5G Coexistence Testbed~\cite{nist5g} & Boulder, CO & \checkmark & NK  & $\times$ & $\times$ & $\times$ & 5G coexistence testing & Indoor & Public* \\
\hline
NIST NBIT Testbed~\cite{nistIntertb}& NK & $\times$ &  & $\times$ & $\times$ & $\times$ &Spectrum sharing  & Indoor & Public*\\
\hline
NITOS~\cite{6932976} &Volos, Greece &$\times$  &\checkmark &$\times$   &$\times$  &$\times$  &Cloud-based Wireless services & Rooftop & Public \\
\hline
Northeastern UAS Chamber~\cite{Northeastern_DroneCage} & Burlington, MA & $\times$ & $\times$ & $\checkmark$ & NK & $\times$ & Drone flights & Drone cage, anechoic chamber & Public*\\
\hline
ORBIT~\cite{raychaudhuri2005overview} & N. Brunswick, NJ & \checkmark  & $\times$ &$\times$   &$\times$  &$\times$  & SDRs & Indoor grid & Public \\
\hline
PNNL 5G Innovation Studio~\cite{PNNL_Testbed} & Richland, WA & $\times$ & $\times$ & $\times$ & $\times$ & $\times$ & Commercial 5G &Indoor & Private\\
\hline
POWDER-RENEW \cite{breen2020powder} & Salt Lake City, UT& \checkmark &\checkmark & $\times$ & $\times$ & \checkmark  & SDRs, massive MIMO & Urban & Public \\
\hline
RELLIS 5G testbed \cite{attTamu} & Bryan, TX & $\times$ & NK & NK & NK &  & 5G (AT\&T) & Outdoor & Public* \\
\hline
Cyber Living Innovation Lab~\cite{cybergmu} &Fairfax, VA & NK &\checkmark  &NK  & NK & $\times$  &5G security, robotics  & Indoor & Public*\\
\hline
SOAR~\cite{SOAR_DroneCage} &Buffalo, NY & $\times$ &  $\times$ & \checkmark  & Partial & $\times$ & Drone flights & Drone cage & Public*  \\
\hline
TIP Community Lab~\cite{tiplabs}& Overland Park, Kansas & NK & \checkmark & $\times$ & $\times$ & $\times$ & O-RAN 5G NR (Sprint) & NK &  Private \\
\hline
UNH Interoperability Lab~\cite{unhinterlab} & Durham, NH & $\times$ & \checkmark & $\times$ & $\times$ & $\times$ & Interoperability testing & Indoor & Public*\\
\hline
Virginia Tech Drone Park~\cite{VT_DroneCage} &Blacksburg, VA & $\times$ &  $\times$ & \checkmark  & Partial & $\times$ & Drone flights & Drone cage & Public*  \\
\hline
\end{tabular}
\label{table:testbeds}
\end{table*}

Even before moving on to discussions of testing requirements specific to Open RAN or UAVs, we can note a few points from  Table~\ref{table:testbeds}.  Naturally, those we were able to survey were largely public-use testbeds, since those are the ones that are most likely to provide information publicly about themselves.  This dovetails with our focus since the interoperability focus of Open RAN implies that for engendering maximum confidence, the testbed facility should be open to anybody that is interested in repeating experiments and verifying results.

Unsurprisingly, there is no testbed on the list that provides full Open RAN as well as UAV support today, even without considering controlled mobility.  Less obviously, we find that the combination of UAV support and controlled mobility is rather rare; only a handful of testbeds on our list provide even partial mobility control in conjunction with UAV support.

Interestingly, we note that a number of testbeds provide emulation support, in keeping with our expectation that this is a key required feature of wireless testbeds.  However, when emulation is considered jointly with mobility control, a non-obvious consideration may be worth mentioning.  For a testbed that provides mobile airborne components, any emulation system must not only emulate the physical RF environment, but also the physics of airflow and aerial navigation, including wind gusts and other disturbing factors (analogous to noise and interference in the RF environment), as well as the dynamics, features, and constraints of a specific UAV. The ability to autonomously navigate one or more UAVs in the 3D space based on RF observations in the environment is also an important capability with various use cases. Furthemore, subtle moves of the UAVs (e.g., a multicopter pitching to move forward) can change the orientation of highly directional RF antennas (especially relevant for mmWave transmissions). With this in mind, it is perhaps unsurprising that the {\em combination} of emulation support and mobility control is quite rare in the extant testbeds.
\subsection{Extant Industry Open RAN Testing Practices}

The facilities listed in Table~\ref{table:testbeds} are largely those focused on system testing, some of which currently already support deploying some particular Open RAN system in part or in full.  Researchers or ecosystem developers may find this sufficient since it is possible for them to test or study their products or innovations in contiguous areas supported by ``some'' Open RAN implementation.  However, vendors, carriers, and other ecosystem players who are involved in the business of actually building or operating a data network as a service need to focus far more deeply on component testing, and (critically for Open RAN) cross-vendor interoperability testing - especially the large swathes of new interoperability modes enabled by Open RAN's disaggregation modes.

Such testing proceeds by identifying Key Performance Indicators (KPIs) of interest, and then measuring them for Devices Under Test (DUT) or System Under Test (SUT) for comparison purposes, as well as possible absolute acceptance criteria.  It would seem a reasonable expectation that an Open RAN system testbed should enable such KPIs to be measured, not just end-to-end, but at interoperation points or interfaces (and for specific O-RAN alliance defined interfaces, including F1/W1/E1/X2/Xn).

However, once one enters the domain of detailed KPIs, there is little standardization of what to measure.  To an extent, the detailed definition of KPIs is part of the specialized knowledge of vendors, operators, and testing service providers that are perceived to provide a competitive advantage, and hence considered confidential.  Because many of the KPIs may be specific to specific vendors, there are also a very large number of them.  Commercial 5G networks test and validate literally thousands of KPIs; the testing regime of  well-known mobile operators actually includes over ten thousand KPIs.  Many KPIs have sub-KPIs and the RF optimization KPIs are substantial.  This will only increase further with the greater use of disaggregation in Open RAN networks.  There are numerous Open RAN interoperability and validation labs today. There are private and public testbeds supported by vendors, consortia, universities, and the government.  Not all labs concentrate on all parts of the toolchain and ecosystem, most focus on specific aspects; validation testing will be greatly dependent on the use case and focus of the lab.  In the Open RAN ecosystem, the RAN Intelligent Controllers (RICs) allow for x-Apps and r-Apps to use the RIC framework as an engine, but with custom functionality.  This implies that every such app can be expected to have a fairly large number of KPIs associated with it depending on its particular functionality. There is the potential for cross-KPIs between the different apps as well.

In light of this, we are forced to go back to fundamentals in recommending KPI capabilities for Open RAN testbeds.  At the highest level of abstraction, there are certain priority KPIs that are foundational for a validation environment, and detailed consideration of many custom KPIs for various operators and vendors (although we are not in a position to list them here) can be seen to trace back to one or the other of these few foundational KPIs:
\begin{itemize}
\item Ability for UE to attach to the network;
\item UE link quality -- uplink and downlink; 
\item UE throughput -- uplink and downlink;
\item Latency;
\item Retainability;
\item Accessibility; and
\item Optimization
\end{itemize}
Each of these KPIs drives multiple other test parameters and features such as performance, load testing, and RF design and optimization.  At this time, practical Open RAN testing in the real world is largely confined to component testing and using KPIs related to the top few items in the above list; in the future, more testing related to the Accessibility and Optimization KPIs is likely to proceed.

\begin{table}[t]
\caption{Example components for an Open RAN validation environment testbed.}
    \def\arraystretch{1}
\begin{tabular}{|p{3.5cm}|p{4.5cm}|}
\hline \rowcolor{LightCyan}
\textbf{Open RAN Components} & \textbf{Test/Evaluation Components} \\ 
\hline
$\bullet$ 5G core access and/or edge & $\bullet$ A Faraday cage / environment  \\
$\bullet$ O-RAN Radios: gNB/eNB (some at controllable UAVs) 
& $\bullet$ 5G signal analyzer – test and validate measurements  \\
$\bullet$ vRAN SW& $\bullet$ RTSA: Real-time spectrum analyzer\\
$\bullet$
 GPS system(s)/Antenna- for synchronization & $\bullet$ Network analyzer-  antenna system and cable measurements \\
$\bullet$
Forward Error Correction (FEC)& $\bullet$ Antenna testing: anechoic chamber- measure patterns \\
$\bullet$ Edge /Server, part of the core network in a box  & $\bullet$ Smaller Shielded enclosures, Faraday cages for individual UAV testing\\
$\bullet$ (Open) RIC platform & $\bullet$ Traffic generator  \\
$\bullet$ rApps, xApps& $\bullet$ Interferers – for testing purposes\\
$\bullet$ UEs (some at controllable UAV for certain use cases)  
& $\bullet$ Various Adapters: need for every type of connector \\
$\bullet$ ToR switch & $\bullet$ Jumper cables \\
$\bullet$ Cell site routers (CSR) & $\bullet$ Attenuators\\
$\bullet$ Acceleration for Open RAN & $\bullet$ Power splitters / power dividers\\
\hline
\end{tabular}
\label{table:Key_ORAN_Testing}
\end{table}

Finally, an Open RAN testbed must include at least one complete reference Open RAN implementation, both to serve as a benchmark for other components to be tested against, and also to enable system tests to proceed for experimenters who wish to innovate in some, but not all, parts of the Open RAN ecosystem.  While Open RAN provides for a multi-vendor environment in building a network from radios, vRAN software, hardware servers, and related software and services, it is important to note that ``open'' does not automatically or necessarily equate to ``interoperable''.  The same need for system integration of multi-vendor Open RAN networks that has driven the need for open test environments must inform the testbed designer in choosing such reference implementations that are actually workable, and hopefully as compliant with O-RAN interface definitions as possible, so as to be broadly compatible with components and devices that testbed clients may bring in the future.  In Table~\ref{table:Key_ORAN_Testing} we have summarized what we perceive to be key high-level components for an Open RAN validation environment testbed.



\subsection{Supporting UAVs in a Testbed}

In its simplest form, any aerial robot (i.e. an airborne device that stays aloft for significant periods of time and is capable of directed motion) can be considered a UAV, but the term is usually reserved for devices that are capable of full (or at least a high degree of) autonomous operation.  
A UAV can therefore exhibit not only primitive autonomous behavior (pre-programmed/way-point trajectory, heat-seeking, collision avoidance, auto-return-to-launch on predetermined conditions such as GPS-lock-loss), but also more complex operations such as computed conditional sensor-driven on-the-fly trajectory control (such as search-and-rescue), participation in coordinated trajectory control locally (platoon or swarm behavior) or globally (such as UTM -- the US Federal Aviation Authority's Unmanned Aircraft System Traffic Management -- or similar), or dynamic self-aware re-tasking (such as degrading mission parameters for safety if battery reserves fall to risky levels).  

In distinguishing between testbed support of UAVs, it is important to realize that a UAV implies close integration of the onboard computing and communication equipment with the vehicle's command and control.  It is helpful to think of two extreme cases as representative of the two classes.  On the one hand, we can mount a computing/communication device (such as an ordinary smartphone) on a UAV.  The UAV's autonomy, trajectory computation, or command and control, remain completely as before.  The coupling between the UAV and the cellphone it carries as a payload is simply mechanical (but may include antenna mounts or high-gain antennas custom-positioned for the UAV, and common power supply).  At the other extreme, the UAV contains only a single computing/communication device, which is capable of being tasked with complex missions (such as air quality analysis, image analysis based search-and-rescue), and also subsumes the trajectory computation (whether autonomous, command-and-control-based, or based on some coordination) for the UAV;  in this case, the vehicle becomes in effect a peripheral of the onboard computer.

First, we consider the task of integrating support in a wireless testbed for UAVs only, used as vehicles for an airborne UE.  This includes the case where the air vehicle has no autonomy and is controlled by a ground-based operator using a handheld or other radio remote control equipment; and even the case where the air vehicle does not have any controlled mobility (such as free-floating balloons) or any mobility at all (such as tethered aerostats or helikites).  The basic challenge for a wireless testbed to support UAVs is posed not only by the fact that they are mobile (which, after all, ground UEs also exhibit, when users walk or drive), but the fact that they
have a widely varied altitude as well as azimuth compared to traditional UEs on the ground. Both spectrum and latency are KPIs of interest for a UE. The front-haul and mid-haul latencies must provide very low latency to maintain system synchronization and function under a varying altitude of the UE, and the spectrum used for communication can significantly affect the achievable coverage and throughput.  A further challenge is that of antenna occlusion, which some UAVs attempt to mitigate by multiple antenna locations around their bodies.  Some UAVs mount antennas on gimbals in an effort to maintain constant directional properties, others allow for servos to allow controlled pointing of antennas.  These challenges are exacerbated by the fact that most base stations, whether commercial or built out of commodity open technology, exhibit their own antenna coverage patterns, which are optimized for ground coverage.  Studies have shown that the consequence of this optimization is the formation of multiple lobes at increasing altitudes, in complex patterns, that cannot be predicted easily as a function of the altitude of the UAV.

\begin{figure}[t]
 \includegraphics[width=0.95\columnwidth]{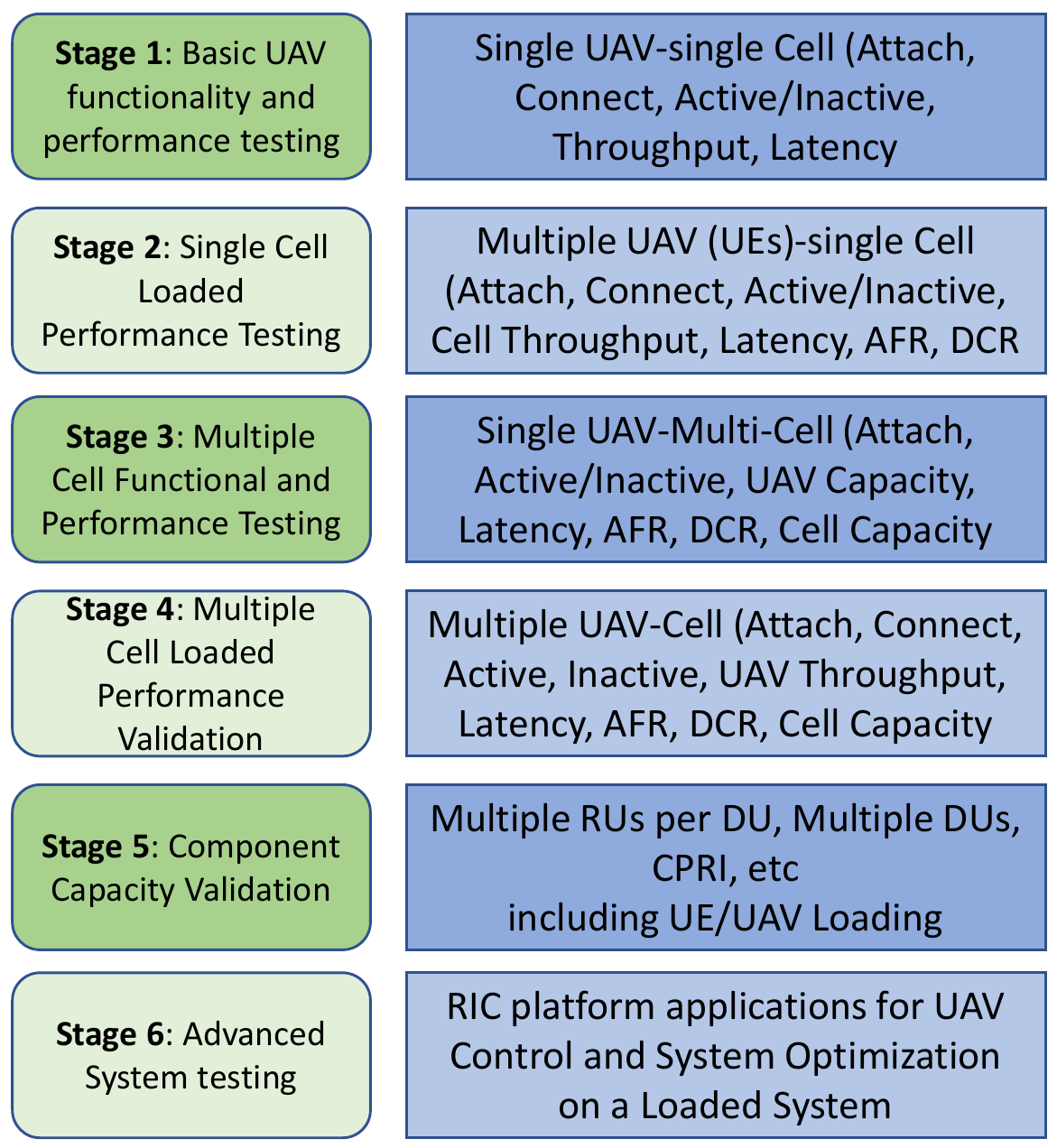}
 \caption{Proposed stages for Open RAN UAV validation.}
 \label{fig:Russ_ORAN}
\end{figure}

The UAV will have to be tested in a controlled environment to ensure the network functions and meet O-RAN specifications. Creating a Faraday environment to do the controlled validation testing will pose challenges compared to traditional Open RAN lab Faraday environments. Then the testing will need to be expanded to an open environment and optimized based on interferers, physical obstacles, and spectrum bands used -- as the propagation and throughput are connected to the spectrum band used for communications. In Fig.~\ref{fig:Russ_ORAN}, we summarize six proposed stages for Open RAN UAV validation. 

While UAVs allow intelligent control of position and trajectory jointly with RAN intelligence (Apps executing at the RICs), the softwarized character of Open RAN also opens up exciting possibilities of allowing the onboard computer to take part in the Open RAN ecosystem in ways other than just as a UE.  We devote the next section to these considerations.

\section{Use Cases for UAVs in Open RAN}  \label{sec:usecase}

\begin{figure*}[th]
	\setlength{\abovecaptionskip}{+0.1cm} 
	\centerline{\includegraphics[scale=0.33]{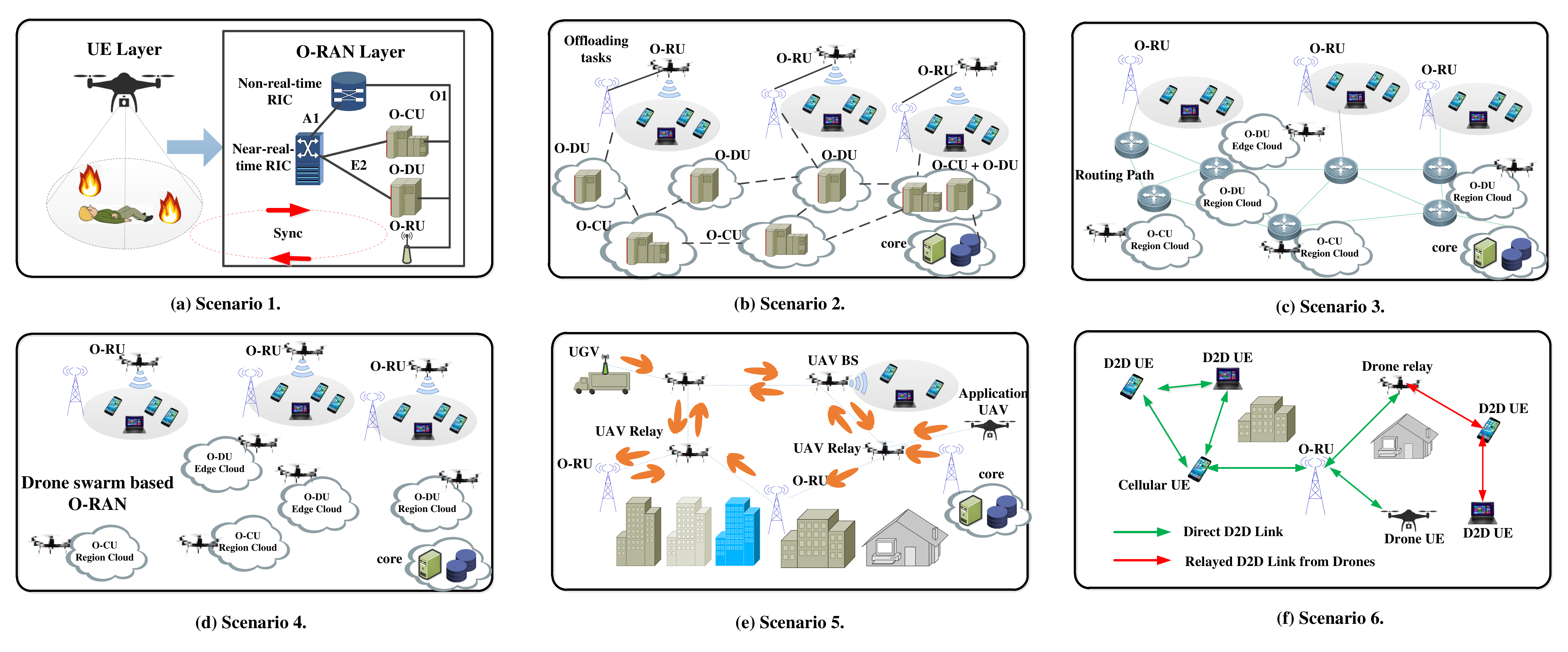}}
	\caption{Use case examples for Open RAN-based air mobility: (a) UAVs as UEs; (b) UAVs as O-RUs; (c) UAVs as O-DUs and O-CUs; (d) UAV swarms in O-RAN; (e) Flying wireless backhaul in O-RAN; (f) D2D communications underlaying UAV-assisted O-RAN.}
	\label{scenario_figure}
\end{figure*}

Considering the aerial controlled mobility and communication among fixed and portable nodes, UAVs will facilitate enhancements to Open RAN with flexible deployments and on-demand, on-time network access. Several use case examples on Open RAN-based air mobility scenarios are provided as follows (see Fig.~\ref{scenario_figure}).

\textit{Scenario 1. UAVs serve as UEs}: This use case focuses on exploring the functionalities of O-RAN RICs for managing and orchestrating network components aimed at 3D critical mission operations (e.g., secure, search and rescue) assisted by UAVs, as they are able to exhibit agile, fast, and autonomous behavior by organizing themselves to exchange information. Considering a scenario involving UAVs connected to an Open RAN ground BS, UAVs as UEs can carry high-resolution cameras and/or sensors, collecting real-time video and transmitting it back to the ground BS, e.g., to be used to identify possible targets of interest through deep neural network object detection model, and in addition report information about application performance to \textit{rApps}. In the meantime, the E2 nodes of O-RAN are responsible for updating UAV control with insights produced by their applications (\textit{xApps} and \textit{rApps}) to support the RAN optimization process.  In this context, Open RAN is able to support the demands of highly dynamic scenarios of critical-mission operations integrated with UAVs due to its flexibility and characteristics of component dissociation.

\textit{Scenario 2. UAVs act as O-RUs}: As described in O-RAN specifications~\cite{ORAN_spec, ORAN_use_case}, UAVs can play a role as O-RUs and process several simple tasks. As the extension, this scenario focuses on the use of UAVs as O-RUs to handle more complicated network tasks, e.g., to quickly deploy an aerial network to assist or extend the terrestrial network where communication and computing resources can move closer to users to meet diverse and stringent 5G application requirements, such as ultra-low latency and ultra-high reliable connectivity. Considering a scenario in which each UAV-BS is equipped with an O-RU to serve ground mobile users, the objective is to optimize the performance of serving offloading tasks via both controlling UAV-BSs to guarantee the quality of communication channels to ground users and efficiently distributing offloading tasks to appropriate Open RAN elements according to the current association. Because of the 3D air mobility capability of UAVs and disaggregation of Open RAN architecture, they may potentially deliver better data offloading capabilities and better resource utilization. 

\textit{Scenario 3. UAVs act as O-DUs and O-CUs}: 1) Using UAVs as O-DUs allows for flexibly hosting RLC/MAC/High-PHY layers based on a lower layer functional split, where UAVs can dynamically connect to multiple O-RUs allowing on-demand resource pooling for virtual baseband functions of high PHY layer, MAC, RLC, and synchronization; 2) using UAVs as O-CUs helps to easily control the operation of multiple O-DUs within/beyond the coverage area, e.g., the radio resource control for flexibly managing the life cycle of the connection, routing or duplication for split bearers, and the service data adaptation for managing the QoS of the traffic flows through autonomous 3D air mobility capability of UAVs.

\textit{Scenario 4. Drone swarm based Open RAN}: This use case envisions multi-role drones without ground facilities that forms an ad-hoc/swarm based Open RAN. Based on \textit{Scenarios 2-3}, we can consider a set of containers to virtualize different O-RAN elements such as O-RUs, O-DUs, and O-CUs deployed in drones and distributed computing nodes of the network. Given these containers with different functions, the objective is to create a robust Open RAN testbed in a swarm of drones towards full decentralization and controlled air mobility. 


\textit{Scenario 5. Flying wireless backhaul in Open RAN}:  Wireless backhaul as an economically sustainable solution has been included by 3GPP as part of the integrated access and backhaul study item~\cite{3GPP_IAB, polese2020integrated} for the 5G NR standard. As an extension in Open RAN architecture, this scenario focuses on building a large-scale, self-organizing network of drones that are connected using a wireless mesh backhaul, which caters to dynamic bandwidth-hungry and latency-sensitive applications. Based on \textit{Scenario 4} with role-specific operations, drones can hover above or close to the O-RU and serve as an airborne last-hop link connecting RAN to the core network. Additionally, they can act as relays between two O-RUs separated by a longer distance to extend coverage forming a multi-hop mesh network for communications and control. Multi-drone backhaul in Open RAN is capable of flexibly adapting itself to cater to highly dynamic applications and events, and easily being scaled up to cover urban scenarios using long-range radios.

\textit{Scenario 6. D2D communications underlaying drone-assisted Open RAN}: Implementation of device-to-device (D2D) communication such as sidelink can be an extension of the network into areas that traditional propagation of the fixed O-RU cannot reach. Particularly, drones can serve as UEs or relays deployed much more swiftly and improve the network throughput performance by dynamically adjusting their locations to provide direct or relayed D2D links to any out-of-coverage users. Additional sidelink capabilities such as multi-hop~\cite{3GPP_R18} and multi-link (in 3GPP Rel.~19) can provide higher resiliency in this mode, especially offering a valuable set of capabilities for mission-critical services such as disaster response rescue and operation.

\textit{Testbed Considerations}: The above poses a rich and variegated set of potential operational scenarios, and it is impractical to attempt to enumerate specific design issues.  Instead, we again propose foundational considerations and hark back to our discussion in Section~\ref{orantb}.  The general capabilities of the testbed that we can identify in order to support such innovative scenarios are:

\begin{itemize}
\item The capability of mobility control of custom air vehicles,
\item The ability to emulate not only the RF environment, but of airflow and UAV flight, and
\item The inclusion of onboard computers, suitable for integration into UAVs, that can support user programming to create software components of the Open RAN ecosystem.
\end{itemize}

\section{AERPAW Testbed Review for Open RAN}

Thus far, we have reflected on general requirements of an Open RAN testbed that is able to integrate UAVs with controlled mobility.  In the remainder of this paper, we take a deep dive into the AERPAW testbed, reviewing it in light of the considerations we have derived above.
We choose AERPAW because we are intimately familiar with it; the authors of this paper include the PIs of the AERPAW project, and key architects and DevOps personnel working on the AERPAW facility.  However, it is also true that AERPAW was conceived and built to support controlled air mobility in a testbed for use by a national community of researchers.  Thus, it is a reasonable facility in which to conduct such a thought exercise of how a fully-featured Open RAN testbed may be built up along the same lines. AERPAW has the foundation for becoming a highly valuable Open RAN UAS test-bed.

AERPAW is the third testbed funded under the PAWR initiative to support advanced and emerging wireless research. It is a multi-year, multi-phase project that started in September 2019 and it is expected to be finalized by 2025. AERPAW experimentation capabilities became generally available with initial set of resources and features in November 2021. Additional platform resources, sample experiments, and experimentation capabilities are expected to be released at the end of Phase-2 (by May 2023) and Phase-3 (by May 2024). 
AERPAW is primarily and essentially a testbed of physical resources, not computing resources. The crucial part of these physical resources are: (i) the RF environment and the airspace that the AERPAW operating areas represent; (ii) the physical equipment (SDRs, commercial RF equipment, UAVs, and UGVs) that AERPAW provides to leverage those environments for experimental studies; and (iii) the expertise (and consequent exemptions) in conducting such studies in compliance with FCC and FAA regulations that AERPAW represents.

Physically, the testbed is hosted at sites in and around the NC State campus in Raleigh, NC. Central to AERPAW's unique characteristic is the availability of UAVs and UGVs in the testbed that can be placed under the direct programmatic control (of trajectories) of the researcher.  In conjunction with the programmable USRPs that are also available for direct programming by the researchers, as well as other real-world, commercial radio equipment, this provides the NextG wireless researcher a facility for research experiments not practicable in any other facility at this time.

\textit{Fixed Nodes, Portable Nodes, and Vehicles: }
At a very high level, the facility includes a number of tower locations (fixed nodes), at each of which some combination of AERPAW programmable SDRs and commercial radio equipment are permanently installed.  The SDRs are controlled by servers, or companion computer (CCs), installed in each location that also represent edge-computing capabilities.  These fixed node locations are distributed over the extensive Lake Wheeler Agricultural Fields of NC State (see Fig.~\ref{fig:LW_deploy}), and some nodes are also installed in the Centennial Campus (see Fig.~\ref{fig:CC_deploy}). The complement of these fixed nodes are AERPAW's portable nodes, also consisting of a computer and SDR(s), but smaller ones so that an AERPAW portable node can be mounted on a UAV/UGV.  The CC on a portable node, an Intel~NUC, also controls the UAV/UGV itself. A smaller version of the portable node that can get carried at the smaller UAV is also available, to do experiments with mobile phones and LoRa sensors that are connected to a LattePanda as the CC.

More information on AERPAW is available at the AERPAW Facility website and User Manual linked therefrom, and previous publications (also listed on the website).  In what follows, we attempt not a comprehensive overview of AERPAW, but rather a review in light of the desirable characteristics we identified above.



\subsection{Span, Scale, Access}

\begin{figure}[!t]
 \begin{subfigure}{0.5\textwidth}
\centering
\includegraphics[width=\columnwidth]{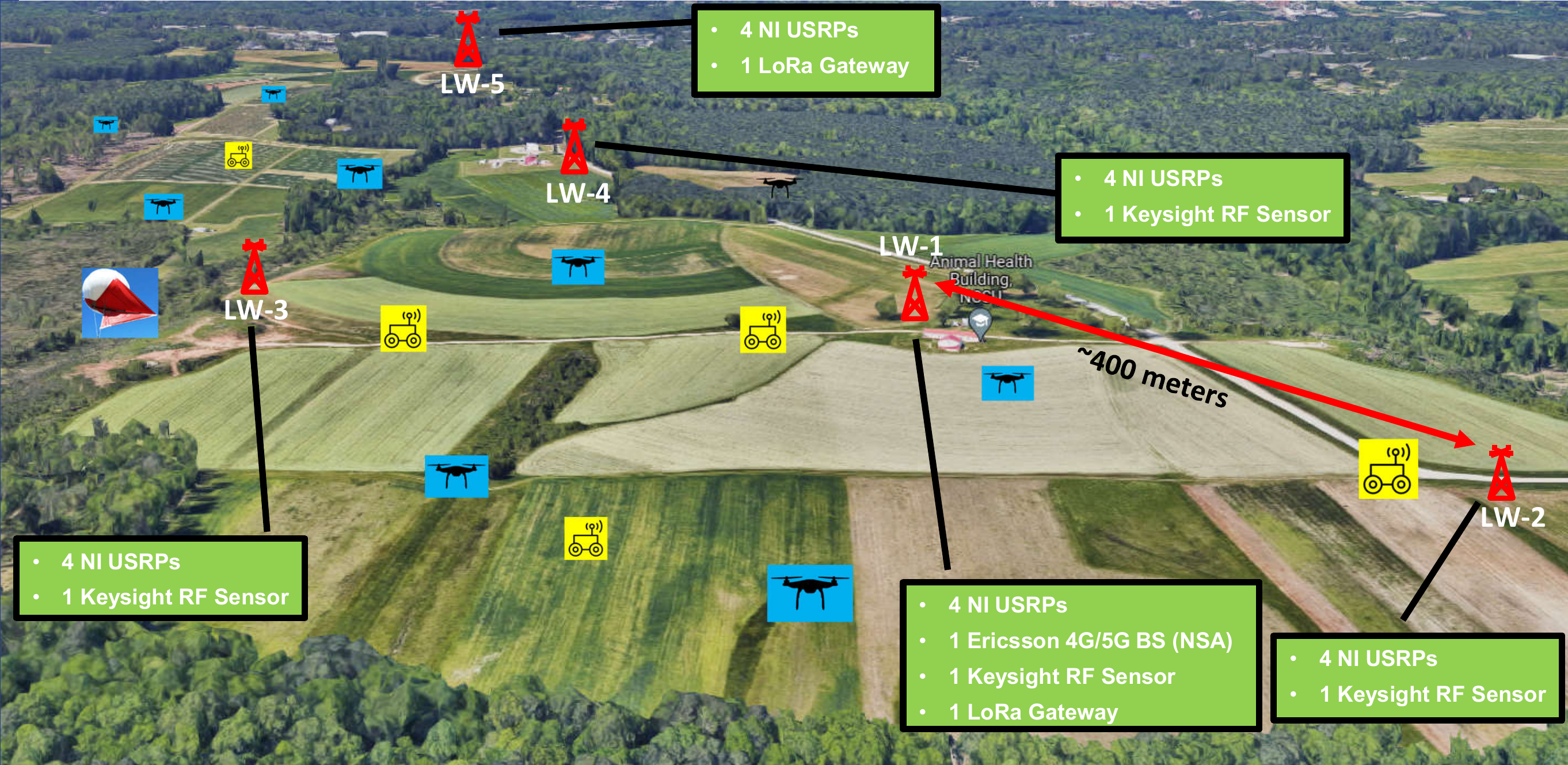}
 \caption{Since Nov. 2021, LW-1 is publicly available for experimentation, and LW-2, LW-3, LW-4, LW5 are expected to be publicly available by May, 2023.}
 \label{fig:LW_deploy}
\end{subfigure}
\hfill\\
\begin{subfigure}{0.5\textwidth}
\centering
 \includegraphics[width=\columnwidth]{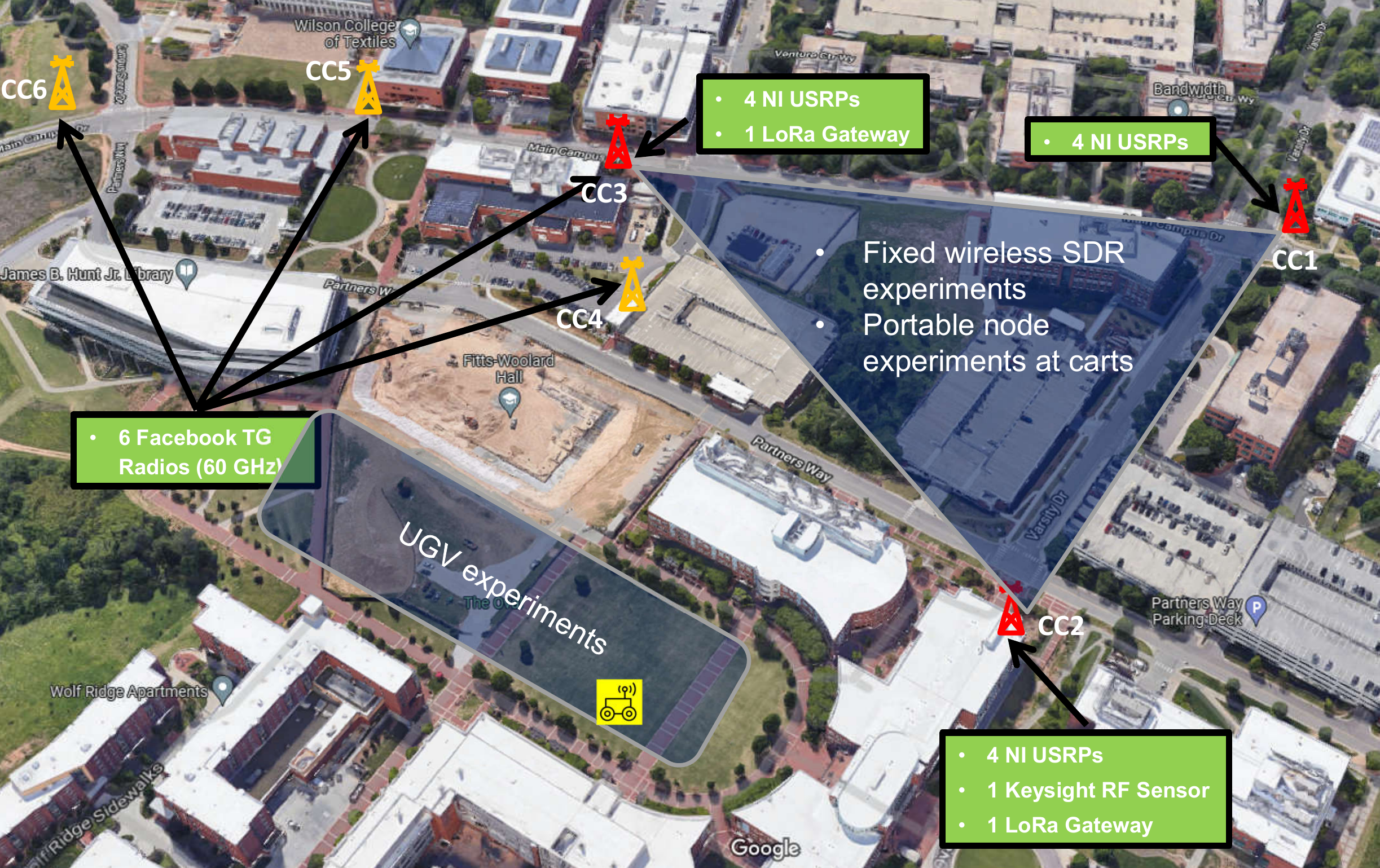}
 \caption{Since Nov. 2022, CC1 and CC2 are publicly available for experimentation, and CC-3 is expected to be publicly available by May, 2023. CC3, CC4, CC5, and CC6 each also has Terragraph radios from Meta operating at 60 GHz.}
 \label{fig:CC_deploy}
 \end{subfigure}
 \caption{AERPAW fixed node deployments at (a) NC State University Lake Wheeler Field Labs, Raleigh, NC; and (b) NC State University Centennial Campus, Raleigh, NC.}
\end{figure}

Fig.~\ref{fig:LW_deploy} and Fig.~\ref{fig:CC_deploy} show the outdoor deployment footprint of AERPAW's fixed nodes in NC State Lake Wheeler and NC State Centennial Campus, respectively.  The equipment that are expected to be available publicly for experimentation by the end of (AERPAW's Phase-2 (expected May 2023) are also illustrated. Currently, it is possible to experiment with UAVs at Lake Wheeler Field Labs; AERPAW does not currently support UAV operation by experimenters in Centennial Campus but supports UGV operation, and UAV operation will likely become available in the future for experimenters.

This geographical span is reasonable for an Open RAN testbed, even with experiments including UAVs.  However, scale is a different matter.  With nine fixed nodes, six portable nodes, eight programmable UAVs, and some non-programmable commercial radio systems such as an Ericsson base station and five Keysight RF sensors, AERPAW can support a large variety of meaningful advanced wireless research -- including proof-of-concept Open RAN experiments at small scales.  But to support the full gamut of Open RAN testing and Open RAN related research experiments, AERPAW would need to add a large number and variety of commercial or stock UEs, and a larger number of programmable UAVs; a few more programmable fixed and portable nodes would also likely be useful.

In Open RAN, the potential softwarization or virtualization of various system components is a particularly attractive feature for innovators.  This requires allowing experimenters direct programming access to all parts of the facility, and at the highest levels of access.  Managing such access while ensuring the safety and regulatory compliance of the facility is a distinct challenge for any testbed that aspires to achieve this.

On this front, AERPAW is already well positioned, having been designed from the outset as a \textit{batch-mode facility}.  Experimenters develop experiments in a virtual environment and submit experiments for execution on the physical testbed once development is complete. AERPAW Operations personnel (Ops) then execute these submitted experiments in the physical testbed environment and collect the output of the experiments as designed by the Experimenters, which are available for Experimenters to view and analyze back in the virtual environment.

\begin{figure}[t]
 \begin{subfigure}{0.5\textwidth}
\centering
\includegraphics[width=\columnwidth]{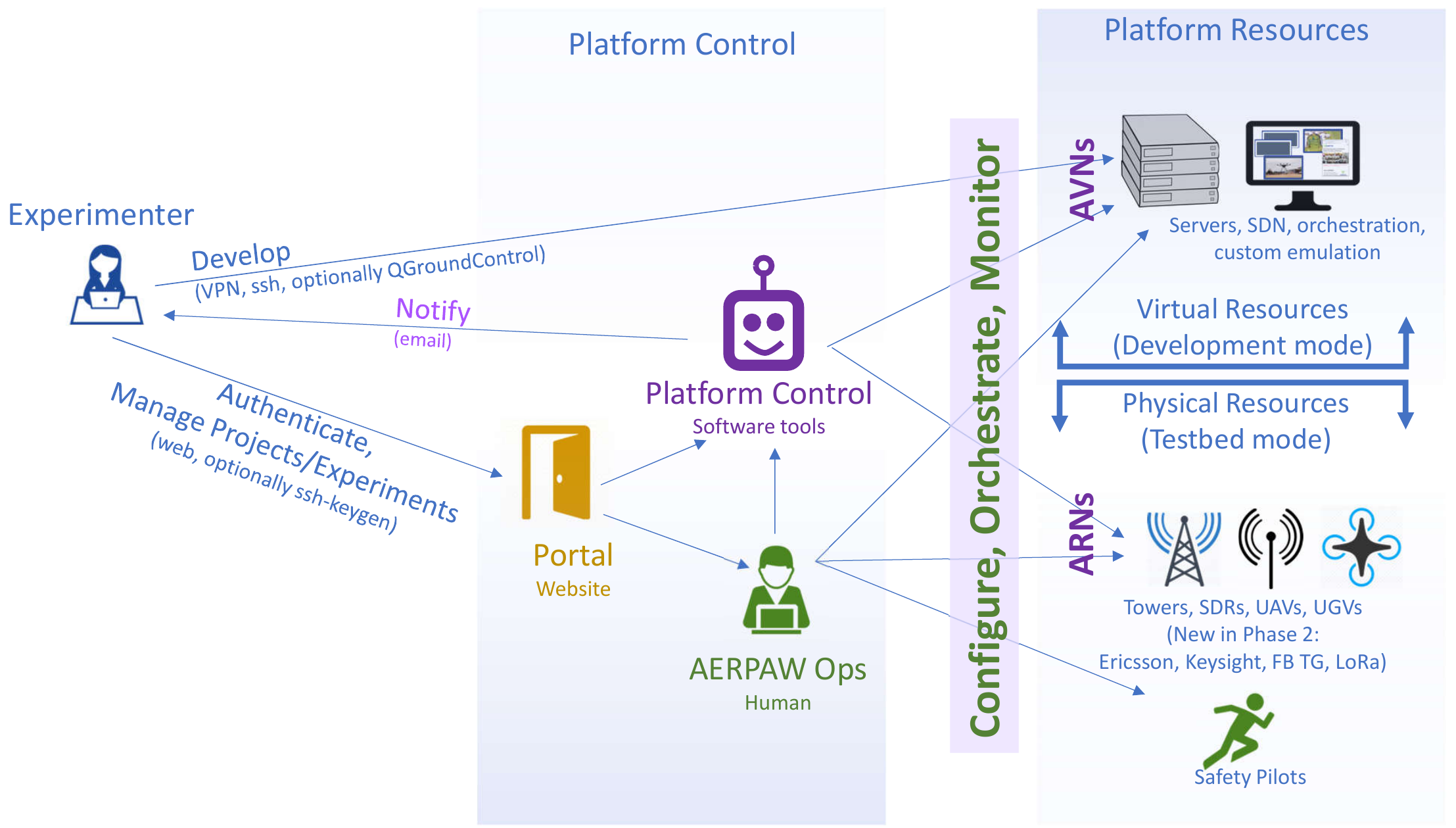}
 \caption{Interaction of an AERPAW experimenter with platform control and platform resources (development mode and testbed mode).}
 \label{fig:workflow1}
\end{subfigure}
\hfill\\
\begin{subfigure}{0.5\textwidth}
\centering
 \includegraphics[width=\columnwidth]{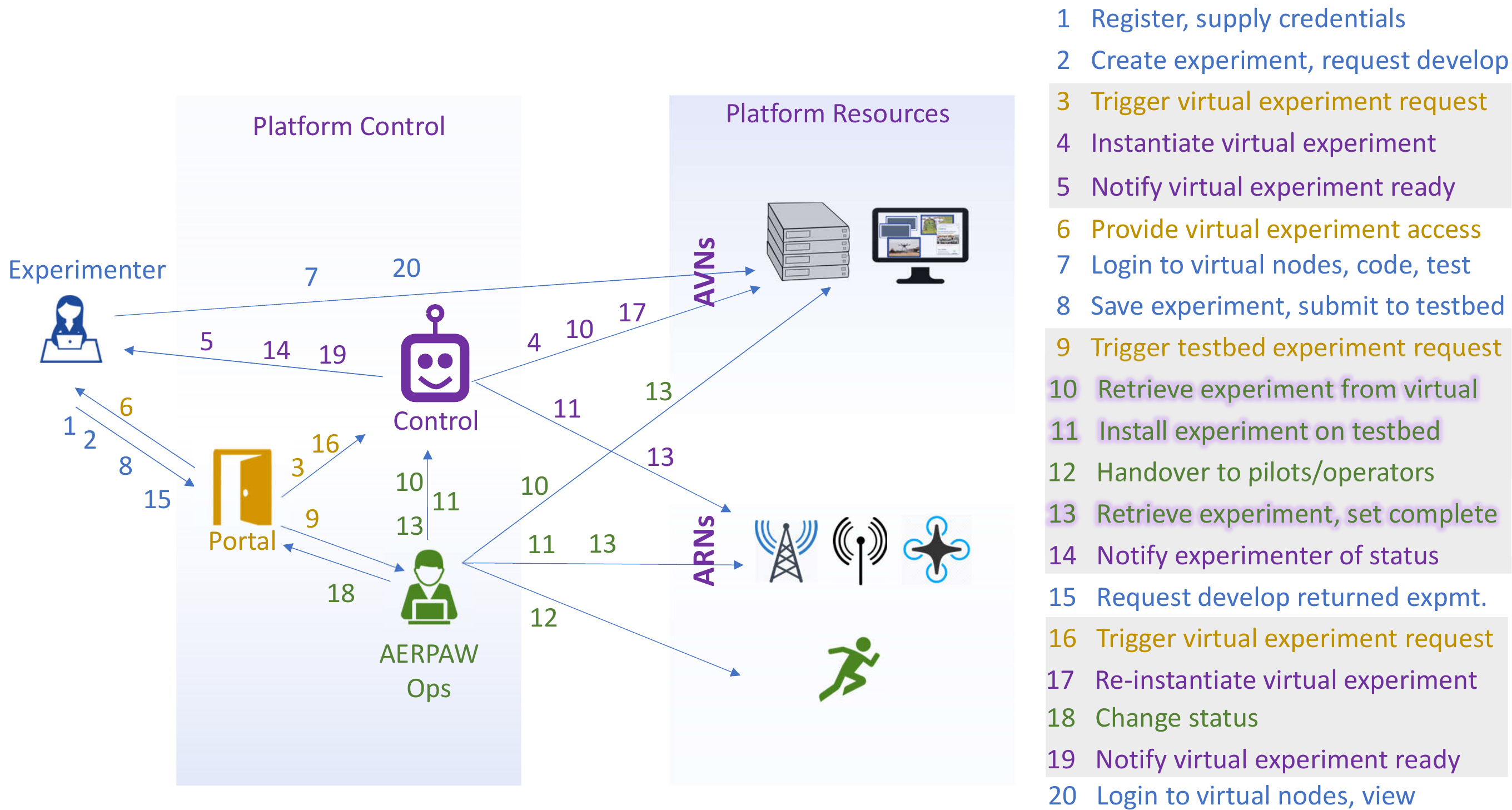}
 \caption{Steps for carrying out an experiment in AERPAW..}
 \label{fig:workflow2}
 \end{subfigure}
 \caption{Experiment workflow for users of AERPAW.}\label{fig:workflow}
\end{figure}

This is not an arbitrarily decided constraint, but a considered architectural choice. In operating a facility with programmable radios and programmable air vehicles, we are obligated to make, and uphold, certain guarantees to the FCC and FAA. However, we also want to allow Experimenters the ability to program those radios and air vehicles, ideally without needing to become fully conversant with FCC and FAA regulation details, obtain exemptions, or expertise in techniques for ensuring compliance.
Batch mode operation allows us to interpose critical filters and monitors into the Experiment code execution flow that allow us to guarantee safe and compliant operation. It is one of the most valuable features of the AERPAW platform that we assume this guarantee ourselves, rather than passing on the responsibility for compliant operations (and liability for non-compliance) to the Experimenter.

Figure~\ref{fig:workflow1} and~\ref{fig:workflow2} show the entity relationships in AERPAW, and the experimenter's experiment design workflow.  Experimenters request ``Development Sessions'' in which they program a virtual environment that is programmatically indistinguishable from the computing environment in the physical testbed.  Once completed, they submit such experiments for ``Testbed Execution Sessions''.  The containers housing the experimenter's code is bodily moved to the corresponding nodes in the physical testbed, where they are executed as before, but with additional supervisory containers monitoring for any RF violation or unsafe air-vehicle operating conditions, overriding as necessary.  As an additional line of defense,  human operators in the field are able to issue aborts if the automated system should fail to override.




\subsection{Spectrum and Licenses}

AERPAW supports multiple frequencies for experimentation with its fixed and portable nodes and vehicles. In particular, AERPAW is one of the few FCC Innovation Zones (FCC-IZs) in the United States \cite[\S1.6]{usermanual} with frequency bands that are highlighted in Table~\ref{table:spectrum}. The maximum effective isotropically radiated power (EIRP) limits for fixed stations (FSs) and mobile stations (MSs) are also specified in the table. The FCC-IZ for Lake Wheeler Field Labs site for AERPAW covers an area of approximately 10.5~square miles, while the Centennial Campus FCC-IZ covers an area of approximately 3 square miles. Experimenters can also port their FCC experimental licenses at AERPAW's FCC Innovation Zone. As noted in Table~\ref{table:spectrum}, due to the sensitivities of certain bands and the wide interference footprint of transmissions from an aerial vehicle, FCC does not allow airborne use in certain bands~\cite{FCC_TAC_UAS_Slides}. 

AERPAW currently supports a subset of the frequency bands through additional FCC experimental licenses (FCC Call Sign: WK2XQH~\cite{FCC_Call_Sign_Search}), which are offered to AERPAW's users to carry out over-the-air experiments on the platform. In particular, for SDR experiments, AERPAW has experimental licenses at 3.3-3.55~GHz  and 902-928~MHz, with plans to incorporate this band into the AERPAW FCC-IZ in the future. The experimental licenses for the Ericsson network include 1.7/2.1~GHz for the LTE system and 3.4~GHz for the 5G system. AERPAW also has plans to support generally available experiments using its mmWave SDR framework by the end of Phase-3 using Sivers phased arrays operating at 28~GHz. Spectrum monitoring and passive I/Q data collection experiments can be supported using USRPs and Keysight RF sensors between 100~MHz to 6~GHz. 

A particular spectrum band that is of recent interest to safety and navigation related command-and-control communications for UAVs, and that AERPAW will explore experimental licenses in the future, is 5030-5091~MHz for which FCC recently released a Notice of Proposed Rule Making (NPRM)~\cite{NPRM_5GHz}. Another band that may potentially be used for ensuring vehicle-to-vehicle (V2V) separation with cooperative surveillance in the future for urban air mobility (UAM) scenarios is 1104~MHz (also known as UAT2)~\cite{GAMA_V2V,stouffer2021enabling,stouffer2020reliable}. Additional spectrum bands that are specifically of interest for UAV/UAM scenarios can be found in~\cite{FCC_TAC_UAS_Slides}.

\begin{table}[t]
\caption{AERPAW's FCC Innovation Zone frequencies. Footnotes: 1) Commission rules do not permit airborne use on all or portions of these bands. 2) Any experimental use must be coordinated with authorized users and registered receive-only fixed satellite earth stations. 3) Operations must be coordinated with a spectrum access system administrator. }
    \def\arraystretch{1}
\begin{tabular}{|p{1.5cm}|p{1.3cm}|p{1.5cm}|p{1.1cm}|p{1.1cm}|}
\hline \rowcolor{LightCyan}
\textbf{Frequency Band} & \textbf{Type of Operation} & \textbf{Allocation} & \textbf{FS Max EIRP} & \textbf{MS Max EIRP }\\
\hline
617-634.5 MHz (DL) &  Fixed & Non-federal & 65 & - \\
\hline
663-698 MHz (UL) &  Mobile & Non-federal & - & 20 (dBm) \\
\hline
907.5-912.5 MHz  &  Fixed and Mobile & Shared & 65 (dBm)& 20 (dBm)\\
\hline
1755-1760 MHz (UL) &  Mobile & Shared & - & 20 (dBm)\\
\hline
2155-2160 MHz (DL) &  Fixed & Non-federal & 65 (dBm)& - \\
\hline
2390-2483.5 MHz  &  Fixed and Mobile & Shared & 65 (dBm)& 20 (dBm)\\
\hline
2500-2690 MHz$^{1,2}$ &  Fixed and Mobile & Non-federal & 65 (dBm) & 20 (dBm)\\
\hline
3550-3700 MHz$^{1,2,3}$ &  Fixed and Mobile & Shared & 65 (dBm)& 20 (dBm) \\
\hline
3700-3980 MHz$^{1,2}$  &  Mobile & Non-federal & - & 20 (dBm)\\
\hline
5850-5925 MHz  &  Fixed and Mobile & Shared & 65 (dBm) & 20 (dBm)\\
\hline
5925-7125 MHz$^2$  &  Fixed and Mobile & Non-federal & 65 (dBm)& 20 (dBm)\\
\hline
27.5-28.35 GHz  &  Fixed and Mobile & Non-federal & 65 (dBm)& 20 (dBm)\\
\hline
38.6-40.0 GHz &  Fixed and Mobile & Non-federal & 65 (dBm)& 20 (dBm)\\
\hline

\end{tabular}
\label{table:spectrum}
\end{table}

\subsection{Mobility Control}
\label{sec:vehicleControl}

\begin{figure}[t]
 \includegraphics[width=0.96\columnwidth]{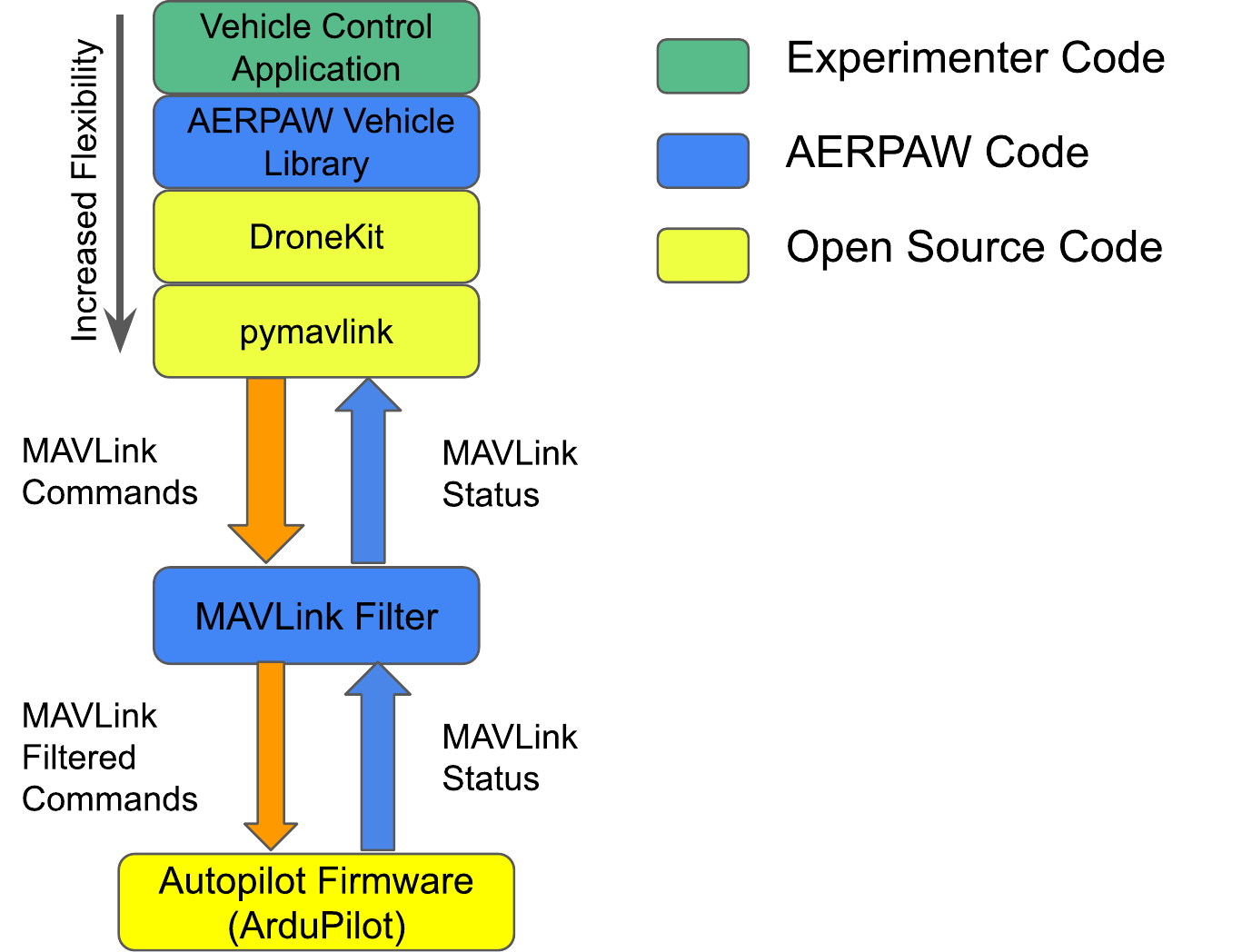}
 \caption{AERPAW vehicle control stack.}
 \label{fig:vehicleControl}
\end{figure}

AERPAW is also, by its original design, already adequate in providing controlled mobility, both for repeatability of experiments and for experimentation with programmatic trajectory control by experimenters; and both for aerial vehicles as well as ground vehicles.
Figure~\ref{fig:vehicleControl} shows the AERPAW vehicle control stack. In AERPAW the main autopilot we support at this time is ArduPilot~\cite{ArduPilot} as it is open source and well-trusted. ArduPilot is supporting MAVLink~\cite{MAVlink} as a communication protocol, and, therefore, all AERPAW vehicle software sends and receives MAVLink commands. For the safety of the testbed and of the AERPAW operators, only a reduced subset of MAVLink commands is allowed to pass through the MAVLink Filter and reach the autopilot. 


Keeping in mind the caveat on the reduced subset of MAVLink commands allowed passing to the autopilot, at one extreme, an experienced AERPAW user can, however, discard the entire stack shown at the top of Fig.~\ref{fig:vehicleControl} and write their own MAVLink application using any other framework they wish (e.g., they could use MAVSDK~\cite{MAVSDK} if they prefer a C++ based library).

However, to smooth the learning curve, we implemented a vehicle library named aerpawlib~\cite{aerpawlib}, which features a finite state machine model, with hooks for vehicle (and/or radio) actions at each state. Several examples are available either to be used as-is or to be modified by experimenters to fit their needs. The most popular example at the moment is the predetermined trajectory sample application, where users specify a series of 3D waypoints to be traversed in order, including choices of the speed and wait times at each waypoint. 

The AERPAW framework also allows the experimenter's programs to take decisions on the fly, thus enabling autonomous applications, such as a radio-based search and rescue (SAR), where the next direction of movement can be chosen based on the current radio measurements.

\begin{figure}[t]
 \includegraphics[width=0.96\columnwidth]{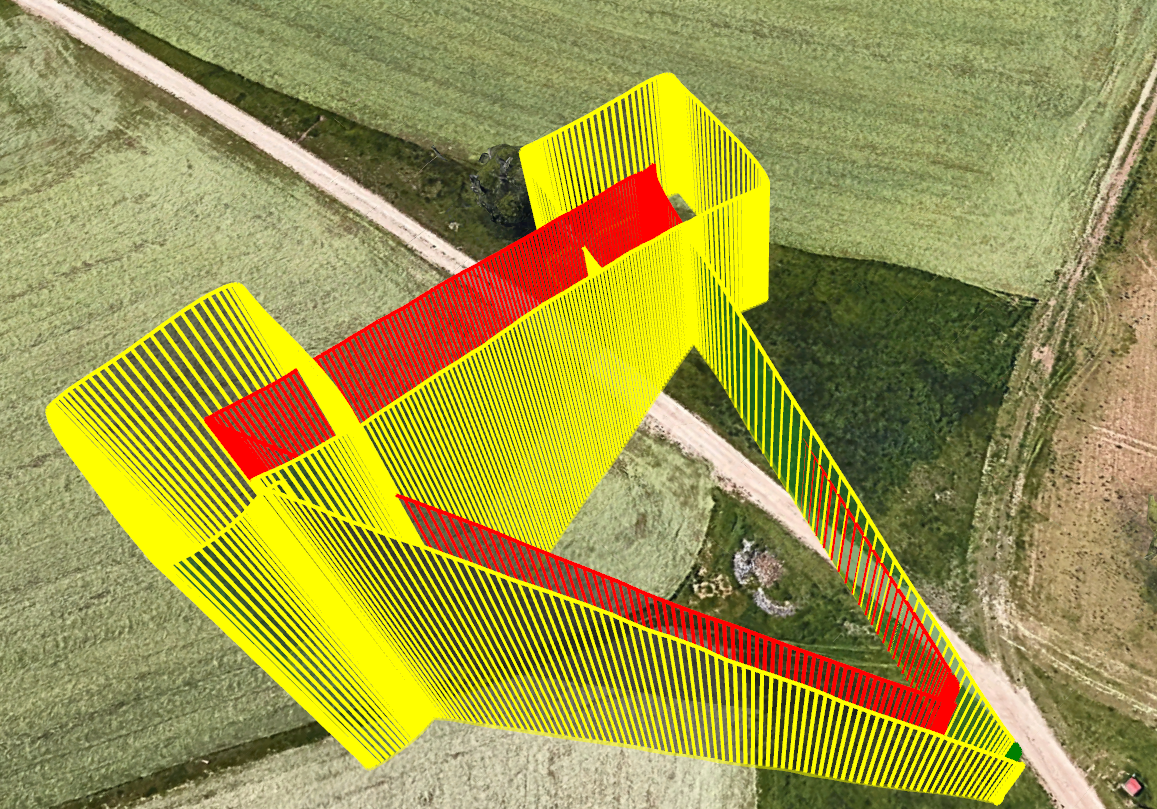}
 \caption{Sample vehicle experiment with two coordinated drones: the tracer (red) goes through a list of waypoints, while the orbiter (yellow) orbits around the tracer while at the waypoint.}
 \label{fig:tango}
\end{figure}

\textit{Autonomous Coordinated Multi-UAV Experiments: }
\label{sec:tango}
An additional feature supported by the application programming library provided by AERPAW is the ability of applications to synchronize the control of multiple vehicles. This is achieved either by using  centralized control (where a coordinator program sends synchronized commands to multiple vehicles), or decentralized applications, (where programs on the companion computer of each of the vehicles coordinate without a centralized conductor). This ability can be leveraged to allow for swarm control.
Fig.~\ref{fig:tango} shows the traces followed by two drones in a coordinated drone experiment, where one drone (the tracer) follows a list of waypoints, while the second drone (the orbiter) shadows the tracer by moving at the same time in the same direction, and upon reaching the target waypoint, it orbits around the tracer once before they both move to the next waypoint. 

This experiment is initially designed and tested in the emulation environment and subsequently executed in the testbed environment. More complicated swarm experiments with a larger number of drones and including communication links with SDRs can be easily carried out using the same workflow. Autonomous decisions can be integrated into the experiment, where the drones can make next waypoint decisions based on the observations of wireless signals. 

Other testbeds can, of course, use alternate methodologies for providing programmatic online trajectory control to experimenters, and repeatability of mobility profiles for experiments.  We have described AERPAW's approach above not to advocate it as the only way, but rather to articulate the level of programmability and repeatability that experimenters should be able to expect from a testbed facility.

\subsection{Emulation Support}

AERPAW has well-articulated emulation support for both RF and air/mobility aspects of experiments.  In the ``Development session'' mentioned earlier, users can prepare their experiments with perfectly repeatable trajectories and wireless propagation. The main goal of providing the emulation environment is to allow users to develop their experiments in a safe and fully repeatable environment.

\begin{figure}[th]
 \begin{subfigure}{0.5\textwidth}
 \includegraphics[width=0.98\columnwidth]{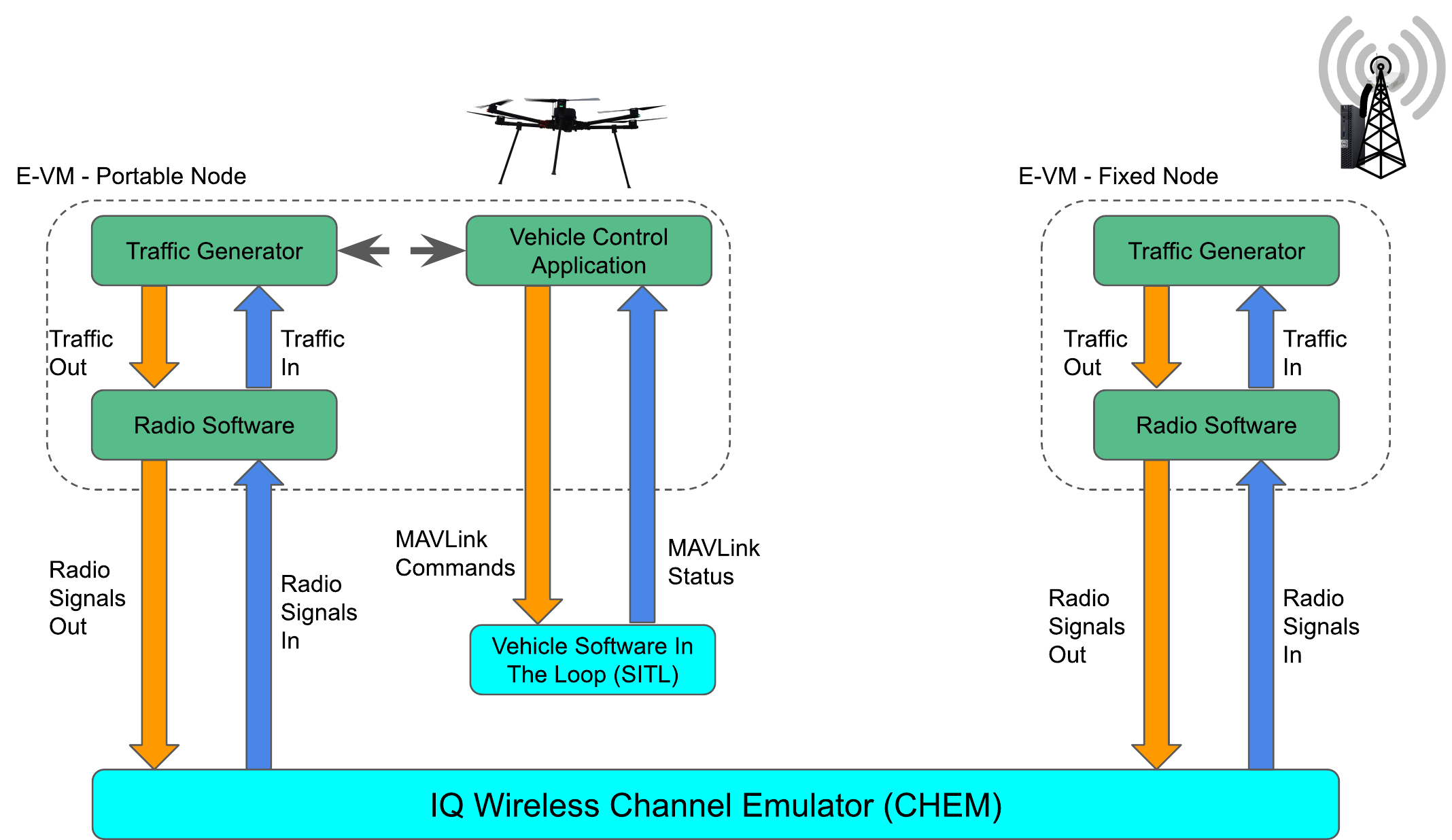}
 \caption{AERPAW emulation environment overview for one mobile node and one fixed node.}
 \label{fig:emulationOverview}
\end{subfigure}
\hfill \\
\begin{subfigure}{0.5\textwidth}
 \includegraphics[width=0.98\columnwidth]{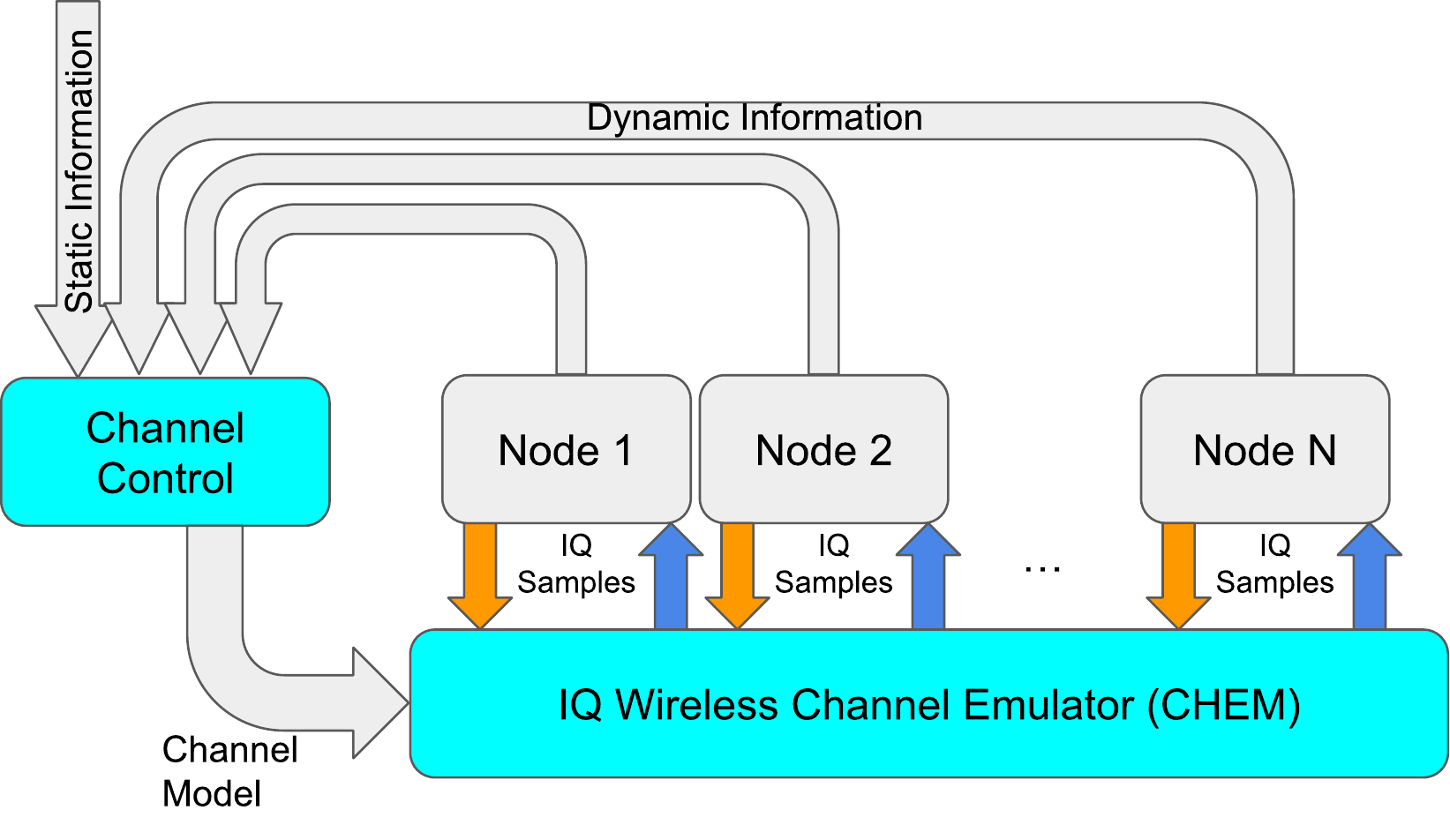}
 \caption{AERPAW wireless channel emulator overview.}
 \label{fig:chem}
\end{subfigure}
\caption{AERPAW emulation environment overview.}\label{fig:chem_overview}
\end{figure}

Fig.~\ref{fig:emulationOverview} depicts an example experiment comprising a portable node on the left and a fixed node on the right while deployed in the emulation environment. In emulation mode, the experimenters' code (encapsulated in the two E-VM, and shown in green in the picture), is running with no modifications in comparison with an experiment in testbed mode. In contrast, in emulation mode, the vehicle and the wireless channel are emulated, thus allowing for a full software emulation, amenable to cloud deployment.

For vehicle emulation, we use an open-source available emulator that has been developed by the ArduPilot community, which features as its main characteristic the use of the {\em same} firmware as the autopilot we use on all our vehicles (at this time, drones, rovers, helikite, and a push-cart). Careful comparisons between the performance of the emulated vehicles and the testbed vehicles show that the vehicle emulator is performing very realistically.

In contrast, for the wireless channel emulator (CHEM), to the best of our knowledge, there is no open-source solution that satisfies all our requirements; therefore, we developed our own solution. Fig.~\ref{fig:chem} shows the main components involved in the CHEM. In general, each radio-enabled node in the testbed is capable of both transmitting and receiving radio signals, which we capture at baseband, IQ level. The IQ samples are sent to the channel emulator, which then ``propagates'' them to the corresponding receivers. The propagation in CHEM is controlled by the channel control module, which dynamically computes a channel matrix based on both dynamic information (e.g., the current mobile node positions and orientations), as well as static information (e.g., position of the fixed nodes, antenna patterns, transmitter gains, etc.). 

The CHEM supports several features, including free space and two-ray ground propagation models, two noise models, MIMO channels, up to 100~MHz of instantaneous bandwidth, multi-rate processing, different antenna patterns, multiple frequencies, and, importantly for efficiency, suppressing silences for bursty traffic.

Once again, we have described AERPAW's approach above not to advocate it as the only way, but to articulate the level of emulation support we find required for an Open RAN testbed.  Regarding AERPAW itself, while it has a good base from which to provide emulation support for Open RAN experiments, it would remain a non-trivial task to develop/procure and incorporate the large volume of software modules that would be required to be integrated into this framework in order to provide emulation support for a comprehensive complement of Open RAN experiments.  In the next section, we return to this topic briefly.


\begin{table}[!t]
\caption{AERPAW example experiments with SDRs.}
    \def\arraystretch{1}
\begin{tabular}{|p{1.3cm}|p{2cm}|p{4.2cm}|}
\hline \rowcolor{LightCyan}
\textbf{Software} & \textbf{Sample Experiment} & \textbf{Comments }\\
\hline
\multirow{6}{1.4cm}{\textbf{srsRAN}}               &  SE1: Multi-node LTE SISO &   Complete end-to-end LTE network with multiple srsUE, and one srsENB and srsEPC\\
\cline{2-3} &SE2: LTE Cell Scan &  Search for LTE cells and capture key parameters of interest  \\
\cline{2-3} &SE3: Two-Node LTE MIMO & Complete end-to-end 2x2 MIMO LTE network, using srsUE with srsENB and srsEPC \\
\cline{2-3} &SE4: Multi-Node IoT &  Basic NB-IoT signalling between the eNB and UE nodes \\
\cline{2-3} &SE5: LTE Handover & Complete end-to-end LTE network with S1 handover, using srsUE with srsENB and open5GS \\
\cline{2-3} &SE6: Single-Node 5G SA & Complete end-to-end 5G SA network, using srsUE with srsENB and open5GS \\
\hline
\multirow{2}{1.4cm}{\textbf{OAI}}               & OE1: Two-Node LTE SISO & Complete end-to-end LTE network, using OAIENB and srsUE \\
\cline{2-3} & OE2: Single-Node 5G SA & complete end-to-end LTE network, using OAIGNB and srsUE \\
\hline
\multirow{3}{1.4cm}{\textbf{GNU Radio}}               & GE1: OFDM TX-RX &  Send and receive data using an OFDM waveform \\
\cline{2-3} &GE2: Channel Sounder &  Pseudo-random sequence of bits are transmitted/received  for channel sounding\\
\cline{2-3} &GE3: LoRa PHY TX0RX & LoRa transceiver with all the necessary receiver components \\
\hline
\multirow{2}{1.4cm}{\textbf{UHD Python-API}}               & UHD1: Spectrum Monitoring & Sweep based spectrum monitoring between 87 MHz and 6 GHz \\
\cline{2-3} &UHD2: IQ Collection &  IQ samples are collected at desired center frequencies with some sampling rate for a specified amount of duration \\
\hline
\end{tabular}
\label{table:SDR_Experiments}
\end{table}

\subsection{Programmability, Radios, Software Stack}

AERPAW does not currently incorporate a full reference O-RAN implementation, although some component parts exist.  The edge-cloud model of companion computers at every AERPAW Radio Node (including both fixed and portable nodes) allows for an easy transition into Open RAN softwarized radio modules, as such modules become available and integrated into the testbed.

The Software Defined Radios of AERPAW represent a potential strength in a possible transition path to full Open RAN support since experimenting with evolving or innovative radio protocols is reduced to an exercise of software development and integration.

AERPAW team provides a variety of SDR sample experiments for experimenters to work with using open-source software and USRP SDRs from NI. Any AERPAW user can start with one of these experiments and develop their code further to research e.g. different protocols and waveforms. 
AERPAW presently supports four different sets of open-source software for SDR experiments: srsRAN~\cite[\S4.1.1]{usermanual}, OpenAirInterface~\cite[\S4.1.2]{usermanual}, GNURadio~\cite[\S4.1.3]{usermanual}, and Python scripts~\cite[\S4.1.4]{usermanual}. A variety of sample experiments are provided in AERPAW's user manual for each case under Section~4.1~\cite[\S4.1]{usermanual}. 

In Table~\ref{table:SDR_Experiments}, we provide a list of SDR sample experiments that are currently available or to be available by the end of AERPAW's Phase-2 (May 2023). An additional set of SDR experiments is expected to be added for general availability by the end of Phase-3 (expected May 2024). All these experiments are tested both in the development environment and the testbed environment of AERPAW. While experimenters can also bring their own software to the platform, AERPAW can not guarantee that they will work smoothly with the existing AERPAW hardware and software, and the development environment. For further details, readers are referred to AERPAW's user manual~\cite[\S4.1]{usermanual}.

\begin{table}[!t]
\caption{AERPAW example experiments with commercial RF hardware.}
    \def\arraystretch{1}
\begin{tabular}{|p{1.2cm}|p{2.1cm}|p{4.2cm}|}
\hline \rowcolor{LightCyan}
\textbf{Software} & \textbf{Sample Experiment} & \textbf{Comments}\\
\hline
\multirow{1}{1.3cm}{\textbf{Ericsson}} &  EE1: 5G Modem RF Logging and Throughput &  Quectel modem logs various KPIs from 4G/5G Ericsson network \\
\hline
\multirow{3}{1.3cm}{\textbf{Keysight RF Sensors}}               & KRSE1: Spectrum monitoring & Monitor and record spectrum up to 6 GHz \\
\cline{2-3} & KRSE2: Signal classification & Classify and detect a variety of signals based on RF signature \\
\cline{2-3} & KRSE3: Signal source tracking & TDOA based localization of a signal source by passive monitoring of its RF signature \\
\hline
\end{tabular}
\label{table:capabilities_commercial}
\end{table}

AERPAW also includes similar prepared experiment profiles for commercial radio equipment available in the testbed (see Table~\ref{table:capabilities_commercial}), but they are relevant in the Open RAN context mainly as potential support equipment, so we do not discuss them further here.

\begin{table*}[!th]
\caption{AERPAW features and capabilities related to Open RAN.}
    \def\arraystretch{1}
\begin{tabular}{|p{2cm}|p{7cm}|p{7.8cm}|}
\hline \rowcolor{LightCyan}
\textbf{Capability} & \textbf{O-RAN Related Components} & \textbf{AERPAW Availability}\\
\hline
\multirow{12}{2cm}{\textbf{Open and Programmable End-to-End Network}}               & Multiple SDRs connected to power and network backhaul & USRPs, Keysight RF sensors \\
\cline{2-3} & Indoor wireless operations in a lab & N/A \\
 \cline{2-3} & Outdoor wireless operations & Rural farm and urban campus\ \\
  \cline{2-3} & Open 5G mobile cores & Open5GS \\
\cline{2-3} & Open fronthaul interface for testing open RUs & Not currently available \\ 
\cline{2-3} & Open source software stacks ready to use with or without additional software development & srsRAN, OAI, GNURadio, I/Q collection with sample experiments~\cite{usermanual}  \\
\cline{2-3} & Open source RIC implementation & Not currently available \\ 
\cline{2-3} & BYOD operation & Yes (on a case-by-case basis)   \\
\cline{2-3} & BYOS operations & Yes (on a case-by-case basis)   \\ 
\cline{2-3} & Bare metal for software installations & Not currently available  \\
\cline{2-3} & Containers for software installations & Yes -- both in emulation and testbed modes  \\ 
\cline{2-3} & Remote access to network resources & Yes during development (emulation) mode, not normally during testbed mode  \\
\hline
\multirow{6}{2cm}{\textbf{End-to-End Network with Commercial Equipment and Swappable Components}}               & Commercial equipment & Ericsson 4G/5G network \\
\cline{2-3} & Indoor wireless operations & N/A \\
 \cline{2-3} & Outdoor wireless operations & Rural farm area \\
\cline{2-3} & Commercial 5G mobile cores & Ericsson NSA core network (Release-15)\\
\cline{2-3} & Includes one or more of a commercial RIC, CU, DU, and RU & Not currently available \\
\cline{2-3} & Open fronthaul interface enabling testing of open RUs to support different physical layers  & Not currently supported \\
\hline
\multirow{4}{2cm}{\textbf{On-site Access to Spectrum}}                  & Unlicensed or ISM band & 900 MHz for aerial communications with SDR front ends \\ 
\cline{2-3} & CBRS spectrum and CBRS SAS features &  N/A\\
\cline{2-3} & Licensed spectrum from a spectrum owner & N/A  \\
\cline{2-3} & Experimental or Innovation Zone licensed spectrum & Yes -- FCC Innovation Zone with 13 bands in 0.6-40~GHz  \cite{usermanual}  \\ 
\hline
\multirow{3}{2cm}{\textbf{Techniques}}                                  & Channel emulation systems & Software emulation available now~\cite{usermanual}, Keysight Propsim (32 ports) channel emulator in the process of integration  \\
\cline{2-3} & Multiple modes of massive MIMO & Not presently available -- mmWave UAV capabilities with 4x4 Sivers phased arrays in development  \\
\cline{2-3}& Emulation capabilities for the RIC, CU, DU, RU, and UE & Presently not available  \\
\hline
\multirow{3}{2cm}{\textbf{Compute Capacity}}                            & One optical hop & Yes  \\
\cline{2-3} & Edge compute & Yes -- Dell 5820 Server at fixed nodes, Intel NUC (i9) at portable nodes carried by AERPAW vehicles \\
\cline{2-3} & Public cloud computing & Not presently supported \\
\hline
\multirow{4}{2cm}{\textbf{Unique Use Case Testing}}                     & Drone support & Multiple different custom drones for different use cases  \\
\cline{2-3} & Rural and urban environment & Yes (autonomous drone experiments available only in rural)   \\
\cline{2-3} & Military base & N/A  \\
\cline{2-3} & Smart agriculture & Deployment in Lake Wheeler agricultural farm of NC State~\cite{usermanual}  \\
\hline
\multirow{5}{2cm}{\textbf{Testing Types}}            & Research and development & Free access by NSF-funded academic researchers, charge-based access for other researchers  \\
\cline{2-3} & Compliance (3GPP, ETSI, O-RAN, etc.) &  3GPP compliant open-source and commercial 4G/5G hardware/software\\
\cline{2-3} & Interoperability & Partial   \\
\cline{2-3} & Security & Partial \\
\cline{2-3} & Performance/stress testing & Partial  \\

\hline
\multirow{4}{2cm}{\textbf{Others}}                                     & Research staff availability & Yes (multiple research associates/students for research support)   \\
\cline{2-3}& Operational staff availability & Yes (multiple research associates/students to support experiments) \\
\cline{2-3}& Wireless certification program &  Not presently supported \\
\cline{2-3}& Established connections to standards/specifications organizations & NextG Alliance, Open Generation Alliance, GUTMA, Linux Foundation InterUSS Platform~\cite{DroneAviationInterop}  \\
\hline
\end{tabular}
\label{table:capabilities}
\end{table*}

\subsection{Summary - Open RAN Related Components of AERPAW}

While AERPAW has not been designed initially as an Open RAN testbed, its open, modular, and flexible design allows possible expanded support for Open RAN use cases as a living lab for UAVs with comparative ease.
The AERPAW team filled out, upon request, a survey in November 2022 developed by the recently established Open RAN working group of the National Spectrum Consortium (NSC)~\cite{NationalSpectrumConsortium}. This survey was shared by NSC members with existing testbed platforms that may potentially support Open RAN experiments in the future. 
In Table~\ref{table:capabilities}, we present a revised version of NSC's Open RAN survey and included comments on AERPAW's features and capabilities that can support Open RAN experiments with controlled aerial mobility. In particular, we highlight open and programmable end-to-end network capabilities as well as commercial 5G equipment deployments in AERPAW, on-site access to wireless spectrum, different experimentation capabilities supported, compute nodes, unique use case testing scenarios, testing types, among other related platform features. 

The information provided in Table~\ref{table:capabilities} relates specifically to the match and extensibility of AERPAW as a meaningful Open RAN testbed for use cases with controlled air mobility.  However, the exercise of preparing this table affords us practical insights into designing and building such an Open RAN testbed, to complement our observations in Section~\ref{requirements}, and we pass these on to the community here.

\section{Representative Results Related to Open RAN   and Controlled Air Mobility}

In this section, we present two early representative experiments from AERPAW that are of relevance for Open RAN experiments. We also elaborate on other possible experiments of relevance to Open RAN that may be supported in AERPAW in the future.

\begin{figure}[t!]
 \includegraphics[width=0.96\columnwidth]{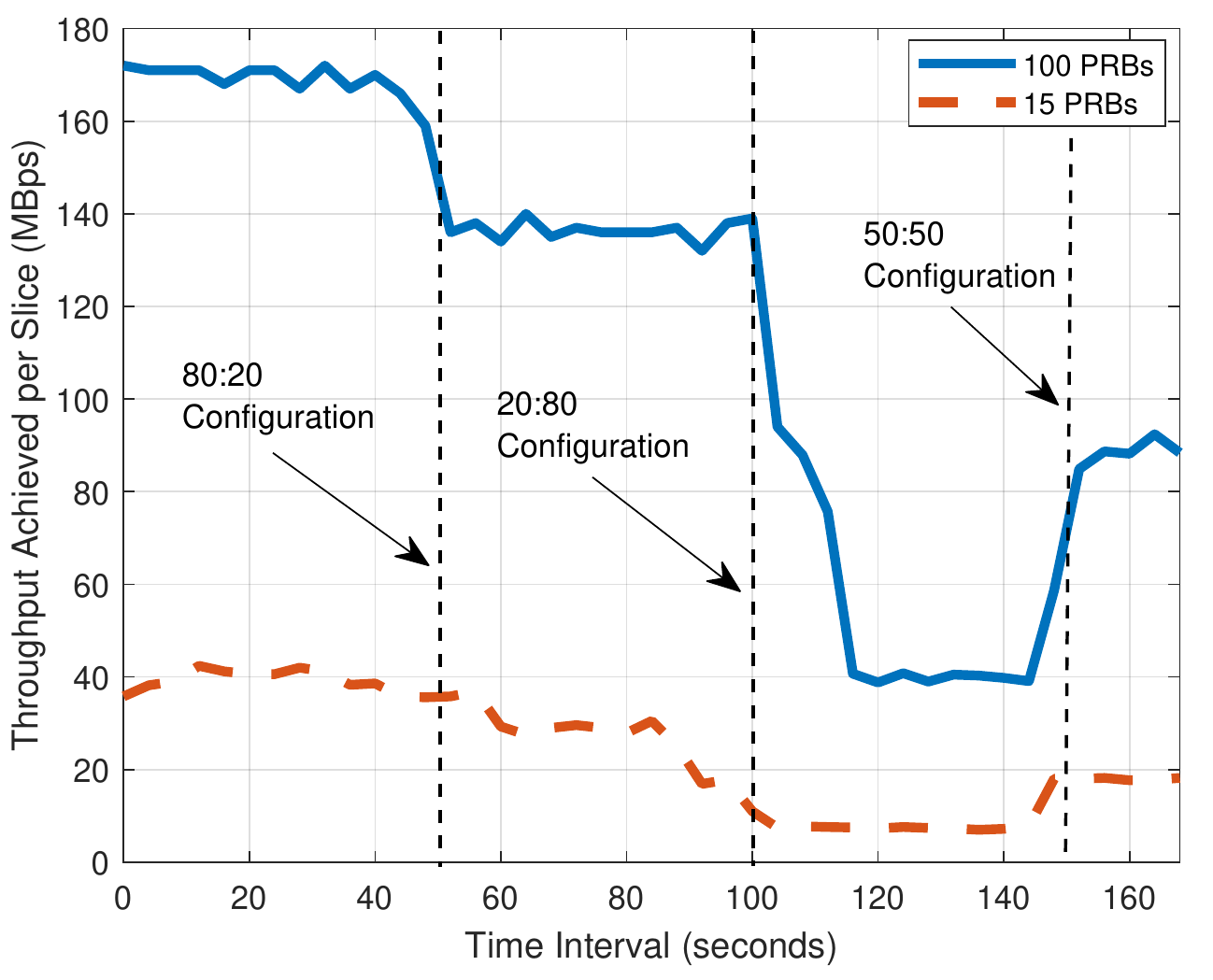}
 \caption{Representative results on O-RAN slicing xApp using srsRAN with two UEs.} 
 \label{fig:ORANSlicing}
\end{figure}

\subsection{RAN Slicing xApp Experiments}

In this section, we provide representative results using the RAN slicing xApp and srsRAN, using the framework by the NSF POWDER Wireless platform~\cite{johnson2022nexran}, executed at the AERPAW testbed.  (Note that these features have not yet been integrated into the AERPAW's development and transition-to-testbed environments; we are exploring integration options at this time).  
The goal is to dynamically create network slices
and observe the effects of slice reconfiguration with a TCP stream on the performance of a UE.
A near Real-time RIC is deployed as part of two separate Kubernetes clusters. Detailed steps are provided in~\cite{OAIC_Github}, we will provide a high-level overview of the architecture. The \emph{RIC cluster} is used for deploying the
platform and applications which are part of the RIC, whereas the \emph{Aux cluster} is used to deploy other auxiliary functions. The RIC Kubernetes cluster installation is done through
configuration scripts and pre-generated helm charts for each of the RIC components. Once the process is done, we created a persistent volume through a
storage class for the influxDB on the RIC platform namespace. Once the RIC platform is deployed, a modified E2 termination is created which has few services enabled to communicate and exchange messages between RIC and E2 Agent~\cite{OAIC_Github}.

Once the Kubernetes clusters are deployed, we can deploy the Near Real-time RIC using a RECIPE file which provides customized parameters for the configuration of a particular deployment group. This Recipe file can be tinkered with if we
want to change any configuration to suit our requirements. Next is the installation of srsRAN components such as srsUE, srsEnB, and srsEPC which use ZeroMQ networking libraries. Since we use ZeroMQ mode, the 4G/5G network can be set up using a single machine that hosts both the RIC and srsRAN
components. Finally, the xAPP is onboarded and deployed on top of the Near real-time RIC and full integration is completed.

Using this setup, we create two network slices in a work-conserving mode and bind two srsUEs to these network slices.
Some representative results are presented in Fig.~\ref{fig:ORANSlicing} for two different bandwidths, which show the throughput of one of the UEs. We configure the slice scheduler in steps to alter the proportionate scheduling in different ways and observe the effects on the TCP stream for the UE \cite{NexRAN_xApp,ORAN_Slicing}. An Iperf server is created on the UE namespace to observe the effects of dynamic RAN slicing and a corresponding Iperf client~\cite{POWDER_ORAN_Profile}.
We create two slices, referred to as \emph{fast} and \emph{slow}, where each slice can be dynamically configured to share the
bandwidth. For the baseline scenario, the full bandwidth of 15 PRBs (100 PRBs) is initially allocated to the unsliced UE which gives a throughput of around 35-40~MBps (170 MBps) as illustrated in Fig.~\ref{fig:ORANSlicing}.

After this, the resources are distributed with the 80:20 configuration among the two UEs. The  results in Fig.~\ref{fig:ORANSlicing} show that the UE's throughput falls to 27~MBps (140~MBps) for this configuration, and when the priorities are inverted between the fast and slow slices to 20:80, the throughput further reduces to 6-7~MBps (40~MBps). Finally, when the priorities are equalized to 50:50 configuration, the throughput increases to 16-17~MBps (70~MBps) for the first UE. The results can be easily extended to a larger number of UEs and more complicated resource configurations.

Our future work includes implementing this same scenario in AERPAW's development and testbed environments with multiple controllable vehicles. The throughput needs and the link qualities of UEs will change dynamically over time as the vehicles move around, and there is a need to have a dynamic slicing mechanism that satisfies the requirements of individual network slices. AERPAW can support development and testing in such dynamic RAN slicing scenarios, first in the emulation environment, and then in the testbed mode with realistic propagation conditions. Programmable mobility with multiple vehicles in both environments and will make it possible to have a testing environment that provides repeatable measurements involving precise mobility control for the UEs, and in some cases, mobile relays and mobile base stations with wireless backhaul.


\begin{figure}[t!]
 \includegraphics[width=0.96\columnwidth]{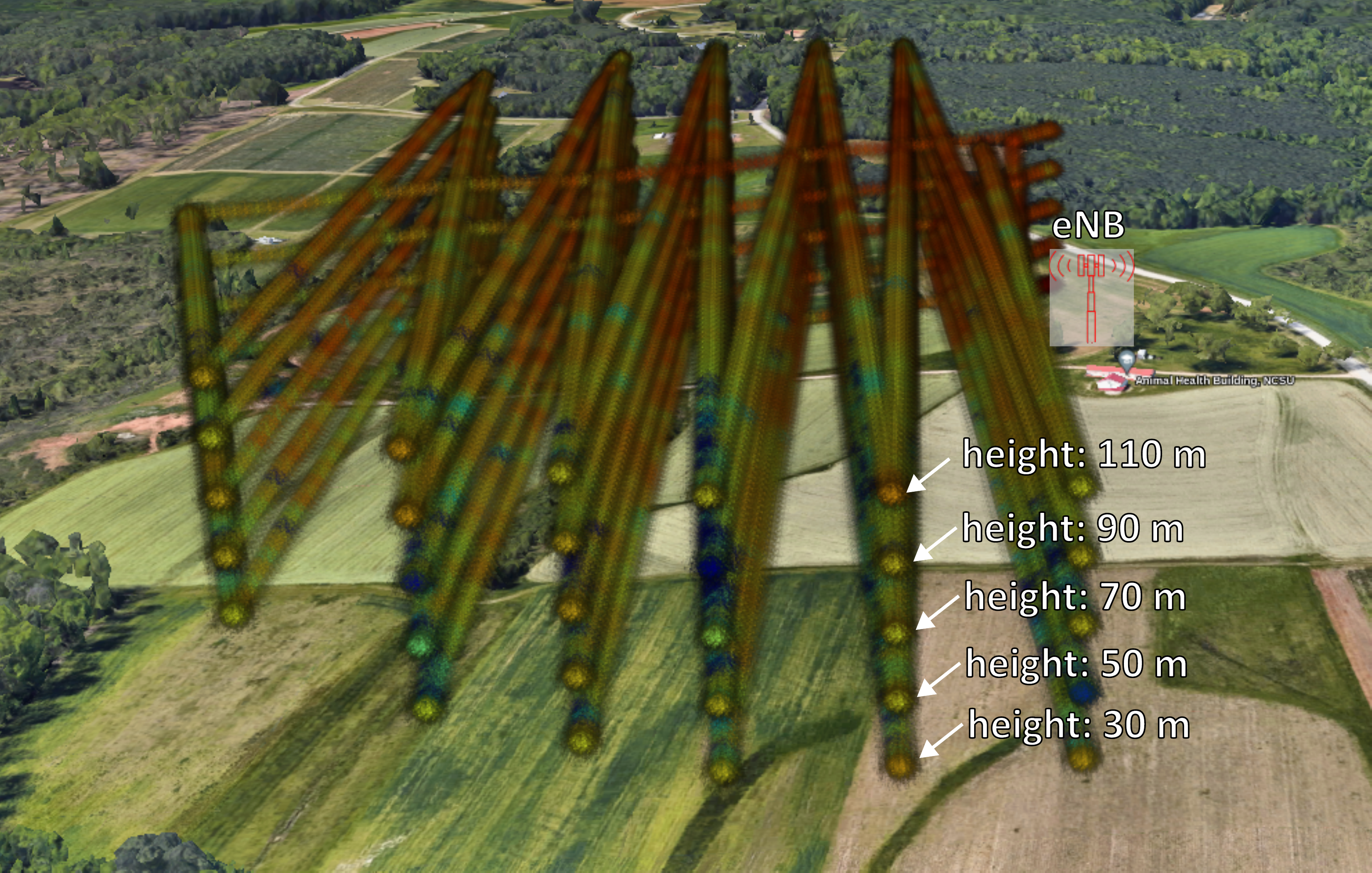}
 \caption{I/Q sample experiments representative results: LTE reference signal received power (RSRP) at five different UAV altitudes.}
 \label{fig:IQ_Experiment}
\end{figure}

\subsection{I/Q Sample Collection Experiments}

In Fig.~\ref{fig:IQ_Experiment}, we provide representative results for the UHD2: IQ collection sample experiment shown in Table~\ref{table:SDR_Experiments}. The UAV is programmed to fly at five different altitudes and the USRP B205mini at the UAV collects IQ samples centered at 3.51~GHz with a sampling rate of 2~MHz. The only signal that can be observed in the spectrogram in the same band is an LTE signal of 1.4~MHz bandwidth, transmitted from a USRP B205 mini that runs srsRAN at our LW1 fixed node. We post-process the collected I/Q samples using Matlab's 4G toolbox, obtain RSRP for each I/Q sample location, and plot the RSRP over the trajectory. Additional details of the measurement setup and representative results are available in~\cite{maeng2022aeriq} using further post-processing with Matlab's 4G toolbox, such as coherence time and coherence bandwidth with respect to the distance between the UAV and the fixed node, kriging interpolation of the received signal across the whole 3D volume, channel estimation, synchronization procedures, among others.

A similar experiment can be carried out to capture I/Q samples and evaluate the KPIs for any Open RAN based 5G system with varying locations of UAVs and UGVs. One or more of the SDR, commercial wireless, or vehicle control sample experiments 
from AERPAW's sample vehicle experiment repository, such as the one illustrated in Figure~\ref{fig:tango} above,
can be used simultaneously with the I/Q sample collection experiment, to collect the raw I/Q data at the finest granularity and post-process them in Matlab's 4G and 5G toolboxes to generate desired KPIs. Such data collected in realistic propagation conditions can be made publicly available to the research community for furthering the research in controlled aerial mobility technologies.

\section{Conclusion}\label{conclusion}

Open RAN expands the capabilities of 5G to support features and functions tied directly to use cases. Disaggregation and virtualization are well suited to UAVs/drones which will continue to grow and become a much greater part of the 5G network from a UE or acting as an O-RU, O-DU, or O-CU component of the network architecture. However, testing and validation are critical to successful integration into 5G and the expansion of Open RAN network capabilities. 

Creating a testbed that supports UAVs poses challenges to meeting all the demands from the physical network to Open RAN interoperability needs. For the UAV market to grow and flourish testing and  validation are necessary. As rules and regulations 
remain volatile in the immediate future,
a UAV Open RAN lab can provide extremely valuable technical results to inform such actions. 

In this paper, we have provided conclusions drawn from our experience and expertise gained from designing AERPAW, a one-of-a-kind public advanced wireless testbed that provides programmable radio and vehicle control in a realistic outdoor area of considerable span, and also reflected on its fit as a possible Open RAN / UAV testbed in future.  We hope these observations may be helpful to the community of designers of other such facilities.

\section*{Acknowledgement}
The authors would like to thank the PAWR Project Office (PPO) and AERPAW project partners including project personnel from Mississippi State University, Wireless Research Center, RENCI, University of South Carolina, and Purdue University, for their contributions to developing the AERPAW infrastructure and for their feedback on this manuscript.   

\bibliographystyle{IEEEtran}
\bibliography{BibFile}











\newpage



\end{document}